\documentclass[12pt]{report}
\usepackage{rotating}
\usepackage[usenames]{color}
\usepackage{overpic}
\usepackage{url}
\usepackage{pst-plot}
\usepackage{amsmath,fullpage,graphicx,caption,float}
\usepackage{amssymb}
\usepackage{array,ragged2e,pst-node,pst-dbicons}
\usepackage{pstricks}
\usepackage{pst-node}
\usepackage{pst-blur}
\usepackage{comment}
\usepackage{color}
\usepackage{subfigure}
\usepackage{lscape}
\usepackage{multirow}
\usepackage{hyperref}
\usepackage{mathtools}
\usepackage{framed}
\usepackage{diagbox}

\usepackage{float}
\restylefloat{table}
\usepackage[margin=25mm]{geometry}

\renewcommand\baselinestretch{1.3}

\usepackage{caption}
\usepackage[english]{babel}
\usepackage{lineno, blindtext}
\usepackage{array}
\usepackage{amsmath}
\usepackage{mdframed}
\usepackage{eso-pic}
\usepackage{lipsum}

\begin{document}

\AddToShipoutPictureBG{%
  \AtPageUpperLeft{%
    \hspace{0.9\paperwidth}%
    \raisebox{-2cm}{%
      \makebox[0pt][r]{CLAS12 Note 2017-001}
}}}%

 \makeatletter
\newenvironment{sqcases}{%
  \matrix@check\sqcases\env@sqcases
}{%
  \endarray\right.%
}
\def\env@sqcases{%
  \let\@ifnextchar\new@ifnextchar
  \left\lbrack
  \def\arraystretch{1.2}%
  \array{@{}l@{\quad}l@{}}%
}
\makeatother

\bibliographystyle{ieeetr}

\pdfcompresslevel=9
\pdfobjcompresslevel=9
\def\be{\begin{eqnarray}}
 \def\ee{\end{eqnarray}}
 \def\ds{\displaystyle}

\newcommand{\mmm}{\mbox{íÜ÷}}
\newcommand{\reff}[1]{(\ref{#1})}
\newcommand{\ra}{\rangle}
\newcommand{\la}{\langle}
\newcommand{\rf}{\}}
\newcommand{\lf}{\{}
\newcommand{\ket}[1]{ | #1 \rangle }



\noindent\begin{minipage}{\textwidth}
\begin{center}
\thispagestyle{empty}
\vspace{0.5cm}
{ \Large{TWOPEG: An Event Generator for Charged Double Pion Electroproduction off Proton}}\\
\vspace{1cm}

{\large Iu. Skorodumina$^{1a, 3a}$, G.V. Fedotov$^{4b, 1b}$, V.D. Burkert$^{2}$, E. Golovach$^{4}$,\\ R.W. Gothe$^{1}$, V. Mokeev$^{2}$}\\[6pt]

\parbox{.86\textwidth}{\centering\small\it
$^1$ Department of Physics and Astronomy, University of South Carolina, 712 Main Street, Columbia, SC 29208, USA.\\
$^2$ Thomas Jefferson National Accelerator Facility, 12000 Jefferson Avenue, Newport News, VA 23606.\\
$^3$ Faculty of Physics, Moscow State University, Moscow, 119991, Russia\\
$^4$ Skobel'tsyn Institute of Nuclear Physics, Moscow State University, Moscow, 119991, Russia\\
E-mail: $^a$ skorodum@jlab.org, $^b$ gleb@jlab.org}\\

\vspace{2cm}
{\bf Abstract}\\[9pt]

\end{center}
{The new event generator TWOPEG for the channel $e p \rightarrow e' p' \pi^{+} \pi^{-}$ has been developed. It uses an advanced method of event generation with weights and employs the five-fold differential structure functions from the recent versions of the JM model fit to all results on charged double pion photo- and electroproduction cross sections from CLAS (both published and preliminary). In the areas covered by measured CLAS data, TWOPEG successfully reproduces the available integrated and single-differential double pion cross sections. To estimate the cross sections in the regions not covered by data, a specialized extrapolation procedure is applied. The EG currently covers a kinematical area in $Q^2$ starting from 0.0005 GeV$^2$ and in $W$ from the reaction threshold up to 4.5 GeV. TWOPEG allows to obtain the cross section values from the generated distributions and simulates radiative effects. The link to the code is provided. TWOPEG has already been used in CLAS data analyses and in PAC proposal preparations and is designed to be used during the CLAS12 era.}


\end{minipage}

\pagebreak


\newpage
\renewcommand{\baselinestretch}{1}\normalsize
\tableofcontents
\setcounter{page}{2}

\newpage
\chapter{Introduction}
\mbox{}\vspace{-\baselineskip}


The GENEV event generator (EG) for double pion electroproduction has been used for many years for the experimental data analysis. GENEV had rather complicated FORTRAN code, and several limitations that prevent its further use during the CLAS12 era. It was based on the differential cross sections from the older JM05 version of the JM model~\cite{Mokeev:2005re},~\cite{Aznauryan:2005tp},~\cite{Ripani:2000va}, which is a reaction model for double pion electroproduction. During the past several years this model has been further developed and significantly improved~\cite{Mokeev:2008iw},~\cite{Mokeev:2012vsa},~\cite{Mokeev:2015lda}. Furthermore,  the GENEV was only applicable up to $W \sim 2$ GeV and for $Q^2 > 0.3$ GeV$^2$, which excludes most of the region that will be under investigation with CLAS12 detector, namely high $W$ and low $Q^2$ (if forward tagger is in use). All these reasons have led to the need to develop a new two pion electroproduction EG TWOPEG, which is the subject of this note.

Section~\ref{sect:two_pi_prod} briefly describes the kinematics of double pion electroproduction off the proton and gives an overview of some details of the experimental cross sections extraction and their subsequent analysis with the reaction model.

TWOPEG uses the method of generation with weights. It means that the EG generates phase space distributions and applies the value of the seven-differential double pion cross section as a weight to each event. This method allows for a significantly more effective and faster generation process, especially in the areas of strong cross section dependencies. All the details of the generation process and obtaining of the final particles four-momenta in the lab frame are given in Sect.~\ref{sect:gen_proc}.

TWOPEG employs the five-fold differential structure functions from the most recent versions of the JM model fit to all results on the charged double pion photo- and electroproduction cross sections from CLAS (both the published and preliminary data~\cite{Ripani:2002ss},~\cite{Fedotov:2008aa},~\cite{Golovach:note}). In the kinematical areas covered by CLAS data, TWOPEG successfully reproduces the available integrated and single-differential double pion cross sections. The quality of the description is illustrated in Sect.~\ref{quality}, where the EG distributions are compared with the available data.

In order to extend the EG to areas not covered by the existing CLAS data, special extrapolation procedures have been applied that included additional  world data on the $W$ dependencies of the integrated double pion photoproduction cross sections~\cite{Wu:2005wf},~\cite{ABBHHM:1968aa}. The new approach allows for the generation of double pion events at extremely low $Q^2$ (starting from 0.0005 GeV$^2$) and high $W$ (up to 4.5 GeV). Details on the corresponding weight evaluation are given in Sect.~\ref{sect:data}.

TWOPEG also makes it possible to obtain the cross section values from the generated distributions. It is helpful to obtain the model cross section value at any point of the kinematical phase space that is covered by the experiment and also to predict it in the uncovered areas. Section~\ref{unfold} contains the details of the unfolding procedure. 

The EG also simulates the radiative effects according to the Mo and Tsai approach ~\cite{Mo:1968cg}. Section~\ref{rad_eff} describes the method of calculating the radiative cross section from the nonradiative one, as well as the method to generate the energy of the radiated photon, which is used for estimating the shift in $W$ and $Q^2$ caused by the radiative effects.

Section~\ref{sect:inp_param} contains the description of the format of input and output files, as well as a short tutorial on how to run the code.

The final Section~\ref{sect:concl} contains the link to the repository, where the TWOPEG code is located and also links to the analyses and proposals, for which the EG has already been used.

\chapter{Double pion electroproduction off the proton}
\label{sect:two_pi_prod}
\mbox{}\vspace{-\baselineskip}

\section{Differential cross sections and kinematical variables in the experimental data analysis}
\label{sect:cr_sect_exp}

\begin{figure}[htp]
\begin{center}
\includegraphics[width=8cm]{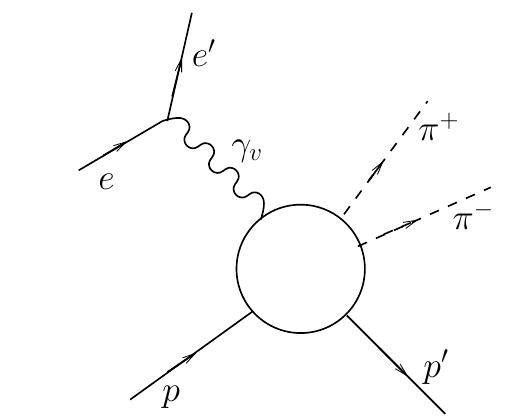}
\caption{\small The process of charged double pion electroproduction off the proton.} \label{fig:2pi_el_prod_scheme}
\end{center}
\end{figure}

The process of charged double pion electroproduction off the free proton is schematically shown in Fig.~\ref{fig:2pi_el_prod_scheme}. The cross section of this exclusive reaction is seven-differential and in the one photon exchange approximation connected with the virtual photoproduction cross section via so-called virtual photon flux $\Gamma_{v}$:

\begin{align} \label{cr_s_el}
 \frac{d^{7}\sigma_{e} }{dWdQ^2d^{5}\tau} = \Gamma_{v}\frac{d^{5}\sigma_{v} }{d^{5}\tau},
\end{align}

where $W = \sqrt{(P_{p}+P_{\gamma_v})^2}$ is the invariant mass of the final hadron system, $Q^2 = -(P_{\gamma_v})^2$ the photon virtuality, $P_{\gamma_v} = P_{e} - P_{e'}$ the four-momentum of the virtual photon, $P_{i}$ the four-momentum of the particle $i$, and $d^{5}\tau$ the differential of the five independent variables of the final $\pi^+\pi^-p$ state.

The virtual photon flux $\Gamma_{v}$ in Eq.~\eqref{cr_s_el} is given by

\begin{equation}
\Gamma_{v}(W,Q^2) =
\frac{\alpha}{4\pi}\frac{1}{E_{beam}^{2}m_{p}^{2}}\frac{W(W^{2}-m_{p}^{2})}
{(1-\varepsilon_{T})Q^{2}} \textrm{ ,}
\label{flux}
\end{equation}

where $\alpha$ is the fine structure constant $\left(1/137\right)$, $m_{p}$ the proton mass, $E_{beam}$ the energy of the incoming electron beam, and $\varepsilon_{T}$ the transverse virtual photon polarization, given by

\begin{equation}
\varepsilon_{T} = \left( 1 + 2\left( 1 +
\frac{\nu^{2}}{Q^{2}} \right)
tan^{2}\left(\frac{\theta_{e'}}{2}\right) \right)^{-1} \textrm{ ,}
\label{eq:polarization}
\end{equation}

where $\nu = E_{beam} - E_{e'}$. $E_{e'}$ and $\theta_{e'}$ are the energy of the scattered electron and its polar angle in the lab frame, respectively. 

\begin{figure}[htp]
\begin{center}
\includegraphics[width=12cm]{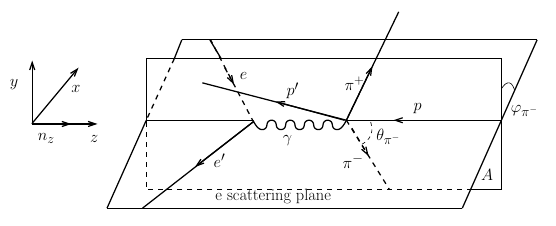}
\caption{\small Polar ($\theta_{\pi^{-}}$) and azimuthal ($\varphi_{\pi^{-}}$) angles of $\pi^{-}$ in the c.m. frame.} \label{fig:cr_sec_thetaphi}
\end{center}
\end{figure}

\begin{figure}[htp]
\begin{center}
\includegraphics[width=12cm]{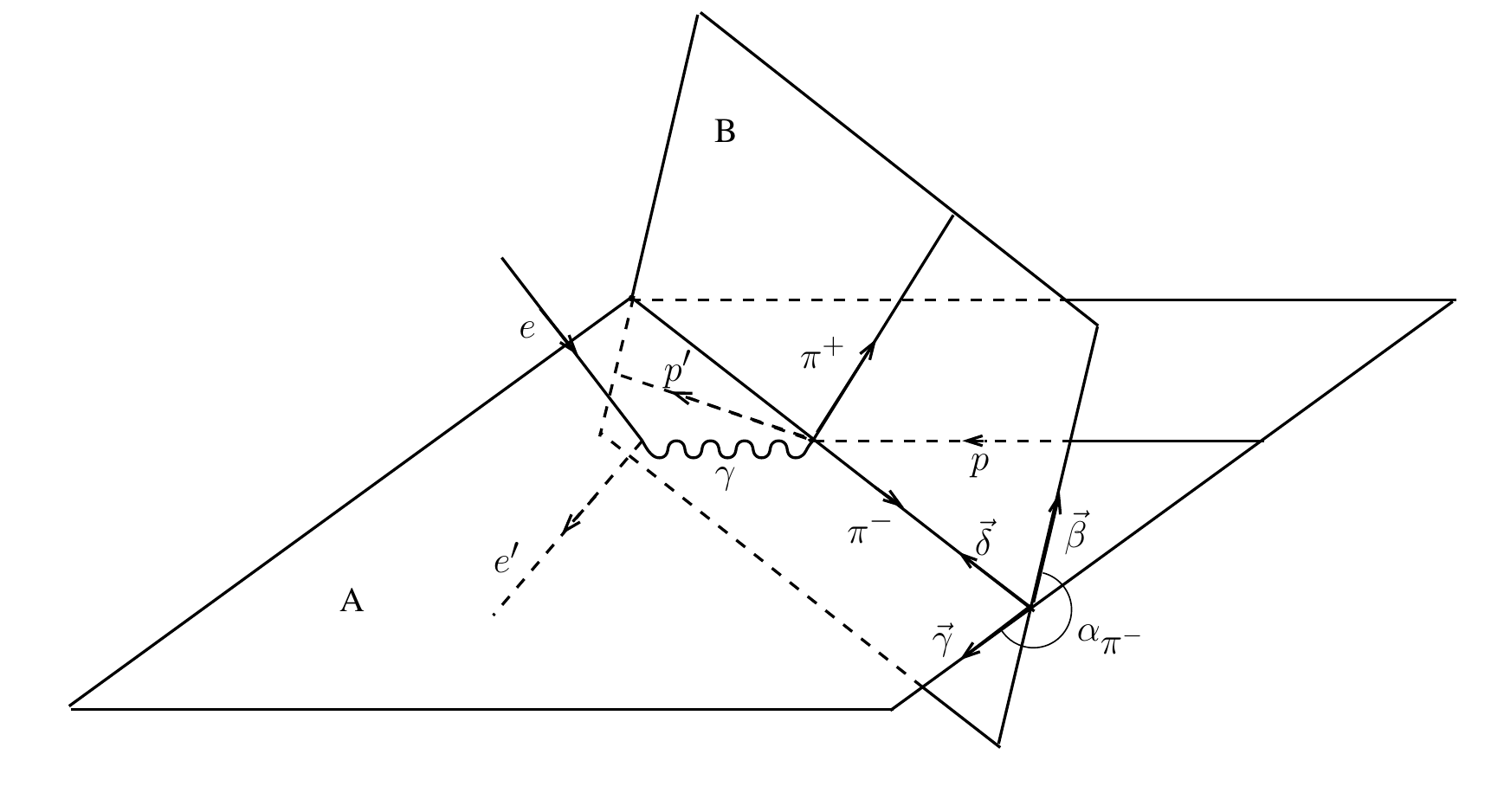}
\caption{\small Definition of the angle $\alpha_{\pi^{-}}$ between two planes: the plane B is defined by the three-momenta of all final hadrons, while the plane A is defined by  the three-momenta of $\pi^{-}$ and initial proton. The definitions of  auxiliary vectors $\vec \beta$, $\vec \gamma$, $\vec \delta$ are given in Appendix~\ref{app_a}.} \label{fig:alpha_2nd_set}
\end{center}
\end{figure}

The conventional choice of five independent variables of the hadron final state is the following:

\begin{itemize}
\item invariant mass of the first pair of the
particles $M_{12}$;
\item invariant mass of the second pair of the
particles $M_{23}$;
\item the first particle's solid angle $\Omega$ = ($\theta$, $\varphi$) (see Fig.~\ref{fig:cr_sec_thetaphi}) and
\item the angle $\alpha$ between two planes: one of
them (plane A) is defined by the three-momenta of
the virtual photon (or initial proton) and the first final hadron, the second
plane (plane B) is defined by the three-momenta of all final hadrons (as shown in  Fig.~\ref{fig:alpha_2nd_set} for the case when $\pi^-$ is chosen as the first particle).
\end{itemize}


These final hadron variables are defined in the center-of-mass frame of the \textit{virtual photon -- initial proton} system.


In the experimental data analysis the cross sections are usually obtained in three sets of variables depending on various assignments for the first, second, and third final hadrons:

\begin{enumerate}
\item $\boldsymbol{1st\; - p',\; 2nd \; - \pi^{+},\;3rd - \pi^{-}}$: $M_{p'\pi^{+}}$, $M_{\pi^{+}\pi^{-}}$, $\theta_{p'}$, $\varphi_{p'}$, $\alpha_{(p,p')(\pi^{+},\pi^{-})}$ (or $\alpha_{p'}$);
\item $\boldsymbol{1st\; - \pi^{-},\; 2nd \; - \pi^{+},\;3rd - p'}$: $M_{\pi^{-}\pi^{+}}$, $M_{\pi^{+}p'}$, $\theta_{\pi^{-}}$, $\varphi_{\pi^{-}}$, $\alpha_{(p\pi^{-})(p'\pi^{+})}$ (or $\alpha_{\pi^{-}}$) and
\item $\boldsymbol{1st\; - \pi^{+},\; 2nd \; - \pi^{-},\;3rd - p'}$: $M_{\pi^{+}\pi^{-}}$, $M_{\pi^{-}p'}$, $\theta_{\pi^{+}}$, $\varphi_{\pi^{+}}$, $\alpha_{(p\pi^{+})(p'\pi^{-})}$ (or $\alpha_{\pi^{+}}$).
\end{enumerate}

Limited statistics of the experimental data does not allow to estimate the five-differential cross section with reasonable accuracy. Therefore, the five-differential hadronic cross sections obtained in each bin in $W$ and $Q^2$ are integrated in order to obtain the single-differential cross sections.

The following set of the single-differential cross sections are usually obtained for the second set of variables:
\begin{equation}
\begin{aligned}
\frac{d\sigma}{dM_{\pi^{+}\pi^{-}}} & =
\int\frac{d^{5}\sigma}{d^{5}\tau}d\tau_{M_{\pi^{+}\pi^{-}}}^{4} & \textrm{with}~~~& 
 d\tau_{M_{\pi^{+}\pi^{-}}}^{4} &\!\!\!\!\! =~~ &
dM_{\pi^{+}p}d\Omega_{\pi^{-}}d\alpha_{\pi^{-}}; \\
\frac{d\sigma}{dM_{\pi^{+}p}} & =
\int\frac{d^{5}\sigma}{d^{5}\tau}d\tau_{M_{\pi^{+}p}}^{4}  & \textrm{with}~~~& 
d\tau_{M_{\pi^{+}p}}^{4} &\!\!\!\!\! =~~ &
dM_{\pi^{+}\pi^{-}}d\Omega_{\pi^{-}}d\alpha_{\pi^{-}}; \\
\frac{d\sigma}{d(-cos\theta_{\pi^{-}})} & =
\int\frac{d^{5}\sigma}{d^{5}\tau}d\tau_{\theta_{\pi^{-}}}^{4} & \textrm{with}~~~& 
d\tau_{\theta_{\pi^{-}}}^{4} &\!\!\!\!\! =~~ &
dM_{\pi^{+}\pi^{-}}dM_{\pi^{+}p}d\varphi_{\pi^{-}}d\alpha_{\pi^{-}}; \\
\frac{d\sigma}{\alpha_{\pi^{-}}} & =
\int\frac{d^{5}\sigma}{d^{5}\tau}d\tau_{\alpha_{\pi^{-}}}^{4} & \textrm{with}~~~& 
d\tau_{\alpha_{\pi^{-}}}^{4} &\!\!\!\!\! =~~ &
dM_{\pi^{+}\pi^{-}}dM_{\pi^{+}p}d\Omega_{\pi^{-}};
\end{aligned}
\label{inegr5diff}
\end{equation}
$$
\text{and\,\,\,} d^{5}\tau = dM_{\pi^{+}\pi^{-}}dM_{\pi^{+}p}d\Omega_{\pi^{-}}d\alpha_{\pi^{-}}. 
$$

For the two other sets of variables the single-differential cross sections can be obtained accordingly.

In the actual cross section calculations the integrals in Eq.~\eqref{inegr5diff} are substituted by the respective sums over the five-dimensional  kinematical grid of hadronic cross sections. 

\section{Differential cross sections in the model analysis}
\label{sect:cr_sect_model}
After the cross sections have been extracted experimentally they are analyzed with a reaction model with the final goal of extracting resonance electrocouplings and revealing and establishing contributions from different reaction subchannels.  

The phenomenological reaction model, which is used for analyzing double pion cross sections, is the so-called JM ("JLab-Moscow State University") model~\cite{Mokeev:2008iw,Mokeev:2012vsa,Mokeev:2015lda}. The model cross section is parameterized in terms of reaction amplitudes and fitted to the experimental single-differential cross sections by adjusting the free model parameters. In this way for each bin in $W$ and $Q^2$ the model produces the five-differential hadronic cross section that can be suitably used in the EG development. 

Let's briefly sketch the framework in which the JM model operates.

For the case when the incident electron beam is unpolarized the virtual photoproduction cross section in Eq.~\eqref{cr_s_el} can be decomposed in the following way (see \cite{Skorodumina:2016pnb} for details):

 \begin{equation}\label{eq:str_fun_decomp}
\begin{aligned}
 \frac{d^{5}\sigma_{v}}{d^{5}\tau }  = \frac{d^{5}\sigma_{T}}{d^{5}\tau }&+\varepsilon _{L}\frac{d^{5}\sigma_{L}}{d^{5}\tau }+ \varepsilon_{T}\left (\frac{d^{5}\sigma_{TT}}{d^{5}\tau } cos\, 2\varphi + \frac{d^{5}\widetilde{\sigma}_{TT}}{d^{5}\tau } sin\, 2\varphi \right )\\
&+\sqrt{2\varepsilon _{L} ( 1+\varepsilon _{T})}\left (\frac{d^{5}\sigma_{TL}}{d^{5}\tau }cos\, \varphi +\frac{d^{5}\widetilde{\sigma}_{TL}}{d^{5}\tau } sin\, \varphi \right ),
\end{aligned}
\end{equation}

where $\varepsilon_{L} = \frac{Q^2}{\nu^2}\varepsilon_{T}$, $\varepsilon_{T}$ is defined by Eq.~\eqref{eq:polarization}, and $\nu$ is the energy of the virtual photon in the lab frame.


The functions $\sigma_{T}, \sigma_{L}, \sigma_{TT}, \widetilde{\sigma}_{TT}, \sigma _{TL}, \widetilde{\sigma}_{TL}$  are known as the structure functions of exclusive meson production. They depend on the variables $W$, $Q^2$, and on all of the kinematic variables of the final state, with the exception of the angle $\varphi$. The $\varphi$ dependence is factorized explicitly by the factors $cos\, 2\varphi$, $sin\, 2\varphi$, $cos\, \varphi$, and $sin\, \varphi$. Consequently any structure function, which is differential in $\varphi$, differs from the same function, which if not differential in $\varphi$, only by the integral factor $2\pi$.

If the cross section given by Eq.~\eqref{eq:str_fun_decomp} is integrated over the angle $\varphi$, only the first (transverse) and second (longitudinal) terms remain.

The functions $\widetilde{\sigma}_{TT}$ and $\widetilde{\sigma}_{TL}$ deserve  special attention: they appear in the case of three-hadron final state ($\pi^+ \pi^- p$) but not in single meson electroproduction and they vanish upon integrating over the angle $\alpha$.

It also needs to be mentioned that the differential cross section in Eq.~\eqref{eq:str_fun_decomp} depends on the beam energy, while the structure functions do not -- the dependence on the beam energy is incorporated into the coefficients in front of them. 

In the limit $Q^2 \rightarrow 0$ that corresponds to the real photoproduction scenario, the coefficients in front of the structure functions force them to vanish, leading $\sigma_{T}$ to be the only remaining term.

As a result of the experimental cross section fitting procedure, the JM model can produce all structure functions from Eq.~\eqref{eq:str_fun_decomp} in the five-dimensional sense. The EG described below is based on these model structure functions.

For the purpose of developing the EG, model structure functions were produced for the second set of variables ($\pi^{-}$ is assumed as the first particle) for those $(W,~Q^2)$ points, where the experimental cross sections are available (see Sect.~\ref{sect:data} for detail). These structure functions are differential in  ($S_{12}$, $S_{23}$, $(-cos\theta_{\pi^{-}})$, $\varphi_{\pi^{-}}$, $\alpha_{\pi^{-}}$), where $S_{12} = M_{12}^{2}$ and $S_{23} = M_{23}^{2}$.

\chapter{The event generation procedure}
\label{sect:gen_proc}
\section{Event generation with weights}
\label{sect:gen_with_weights}



For this EG the method of generation with weights is used. It means that instead of forcing the number of generated events in particular kinematical point to be proportional to the cross section value at this point, the number of generated events is maintained the same everywhere, while each event acquires an individual weight that is equal to that cross section value. In this way the generated events still reproduce the realistic cross section shape if they are summed up with weights.

The weight for each event is calculated based on the model structure functions in the ($W$, $Q^2$) regions, where experimental data exists. In other areas, where no experimental data exists, the weight is estimated based on a special procedure of interpolation and extrapolation of the structure functions from the areas covered with data (see Sect.~\ref{sect:data} for details).

As it is mentioned in Sect.~\ref{sect:cr_sect_model} the model structure functions are differential in the following kinematical variables ($S_{12}$, $S_{23}$, $(-cos\theta_{\pi^{-}})$, $\varphi_{\pi^{-}}$, $\alpha_{\pi^{-}}$) and given for different ($W$, $Q^2$) points for the variable set, where $\pi^{-}$ is assumed to be the first particle. This forces us to use this framework for the event generation. 

Below all subscript indices $1,\;2,\;3$ correspond to the final hadron numbering, which is in the second set of variables the following: $1st\; - \pi^{-},\; 2nd \; - \pi^{+},\;3rd - p'$. The subscript index $h$ corresponds to $hadron$ and in the second set of variables it is $\pi^{-}$.

For each event the values of all kinematical variables $W$, $Q^2$, $S_{12}$, $S_{23}$, $cos \theta_{h}$, $\phi_{h}$, $\alpha_{h}$ are generated randomly~\footnote[1]{If you use ROOT functions, it is convenient to use random number generator class TRandom3. \\
TRandom3 rndm\_name(UInt\_t(((float) rand() / (float)(RAND\_MAX))*4000000000.)); \\
Var =rndm\_name.Uniform(var\_min,var\_max); \\
It is convenient to use separate random generators for each variable that needs to be generated randomly. It is essential because if the same random generator is used for all variables, the generation constrains (discussed in the text under 1. and 2.) may disturb the randomnicity of the generation of some variables.} according to the double pion production phase space. This means that:

\begin{figure}[htp]
\begin{center}
\includegraphics[width=16cm]{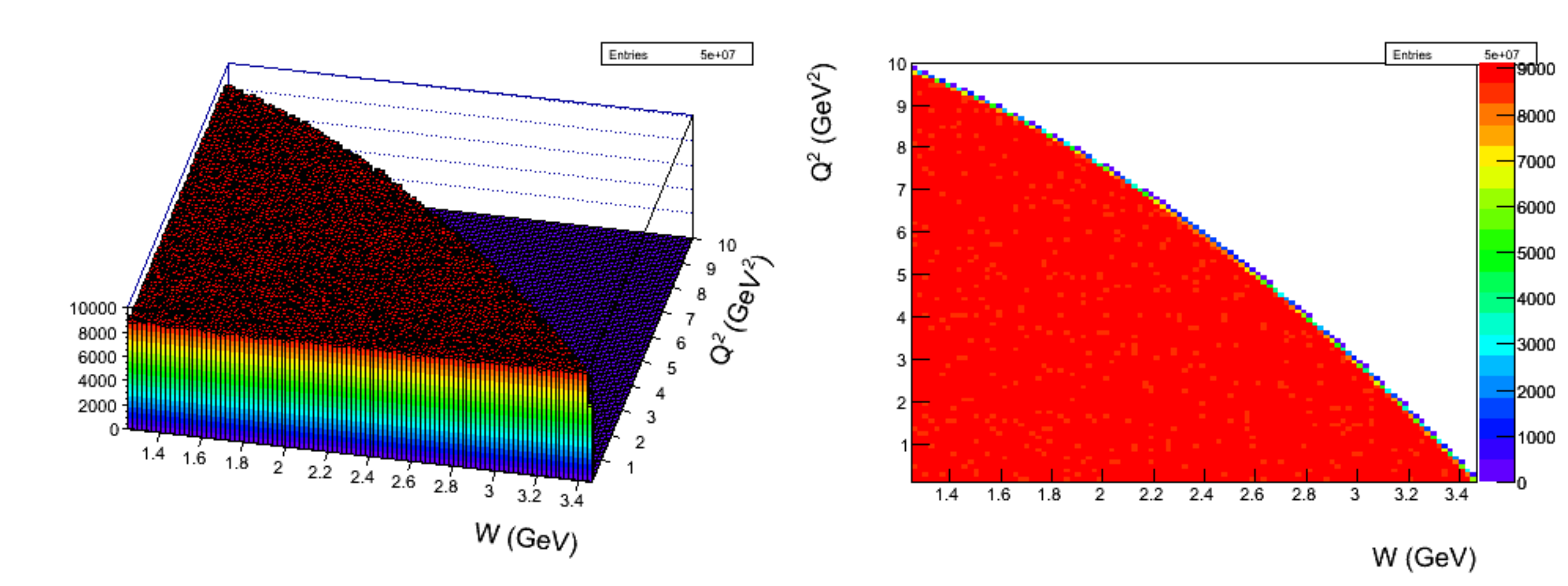}
\caption{\small Generated flat $Q^2$ versus $W$ distribution for $E_{beam} = 6$ GeV. Left plot -- {\it lego2} option, right plot -- {\it  colz} option.   } \label{fig:w_vs_q2_flat}
\end{center}
\end{figure}

\begin{figure}[htp]
\begin{center}
\includegraphics[width=16cm]{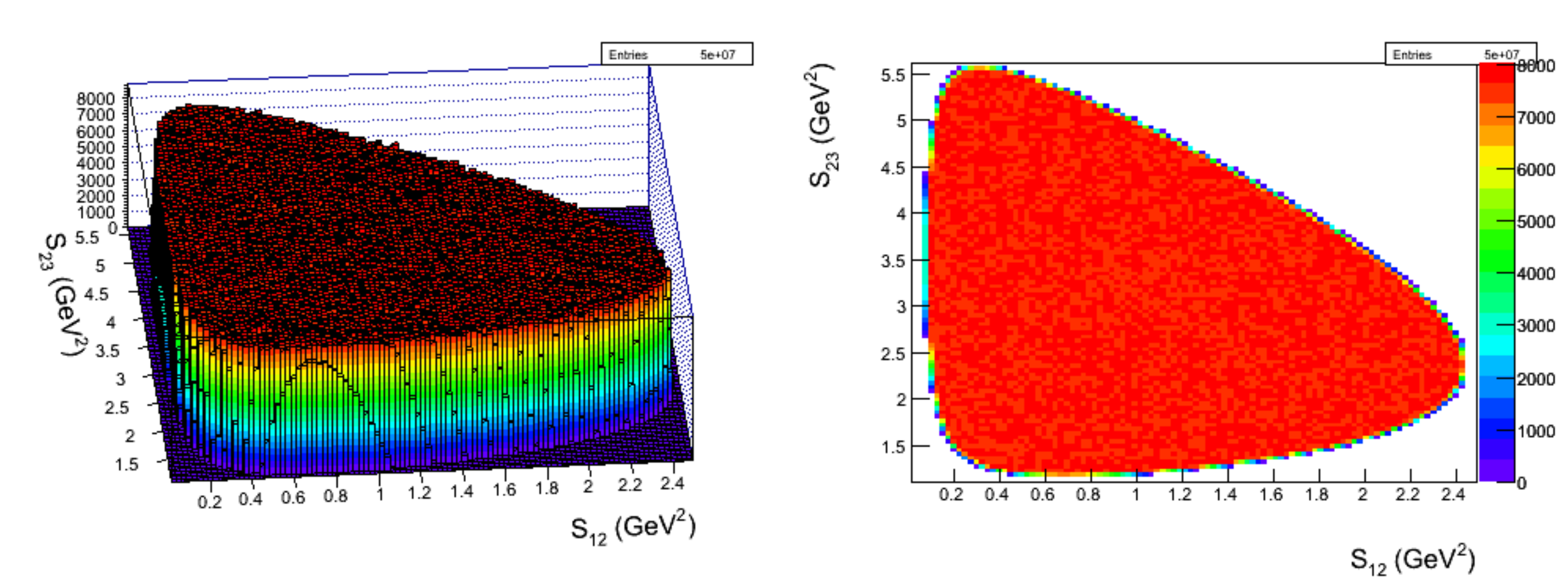}
\caption{\small Generated phase space $S_{23}$ versus $S_{12}$ distribution for  $W = 2.5$ GeV.  Left plot -- {\it lego2} option, right plot -- {\it  colz} option.  } \label{fig:s23_vs_s12_flat}
\end{center}
\end{figure}

\begin{enumerate}
\item $W$ and $Q^2$ are generated flat in two-dimensional sense within the limits $[W_{min},W_{max}]$, $[Q^2_{min},Q^2_{max}]$, which are among the input parameters. Figure~\ref{fig:w_vs_q2_flat} shows an example of a $Q^2$ versus $W$ distribution. The triangle-like shape of this distribution is the consequence of the following constrain: $\nu < E_{beam} - E_{min}$, where $E_{beam}$ is the energy of the incoming electron beam and $E_{min}$ is the minimal energy of the scattered electron, respectively (both of them are defined in the lab frame and given as input parameters). This condition forces the energy of the virtual photon in the lab frame ($\nu$) to have reasonable values.

\item $S_{12}$ and $S_{23}$ are generated flat in two-dimensional sense within the limits $[(m_{1}+m_{2})^{2}, (W-m_{3})^2]$, $[(m_{2}+m_{3})^{2}, (W-m_{1})^2]$, where $W$ is the $W$-value generated in the previous step, $m_{1}$, $m_{2}$, and $m_{3}$ are the masses of the first, second, and third hadron, respectively. Figure~\ref{fig:s23_vs_s12_flat} shows an example of $S_{23}$ versus $S_{12}$ distribution. The shape of this distribution is determined by the condition $B(S_{12},S_{23},W^{2},m_{2}^{2},m_{1}^{2},m_{3}^{2}) <0$, where $B(x,y,z,u,v,w)$ is the Byckling function that is defined in the following way~\cite{Byckling:1971vca}:

\begin{equation}
\begin{split}
B(x,y,z,u,v,w) = &x^{2}y+xy^{2}+z^{2}u+zu^{2}+v^{2}w+vw^{2}+ \\   
&xzw+xuv+yzv+yuw- xy(z+u+v+w)- \\ 
&zu(x+y+v+w)-vw(x+y+z+u).
\label{eq:byckling}
\end{split}
\end{equation}

\item  $cos \theta_{h}$ is generated flat in a one-dimensional sense in the limits $[-1,~1]$. 
\item The angles $\varphi_{h}$ and $\alpha_{h}$ are generated flat in a one-dimensional sense in the limits $[0,~2\pi]$.
\end{enumerate}

Note that those variables must be generated flat in which the model cross section is differential. It is essential for the correct weight propagation.

All generated variables listed above are assumed to be defined in the CMS. 

Furthermore, the azimuthal angle of the scattered electron in the lab frame $\varphi_{e'}$ and its $z$-vertex $Z_{e'}$ are generated flat in a one-dimensional sense in the limits $[0,~2\pi]$ and $[Z^{off}_{targ}  - L_{targ}/2,~Z^{off}_{targ}  + L_{targ}/2]$, respectively, where $Z^{off}_{targ}$ and $L_{targ}$ are among the input parameters.

It also needs to be mentioned that if one uses a weighted EG for the purpose of efficiency evaluation, the statistical uncertainty of the efficiency must be calculated in a slightly different manner compared to the unweighted case. The proper way to calculate the uncertainty is discussed in~\cite{Laforge:1996ts}.

\section{Obtaining particle four-momenta in the lab frame}
\label{sect:cms_to_lab_trans}

The final goal of the event generation process is the following: using the generated values of all kinematic variables described in the previous section, to obtain the four-momenta of all final particles in the lab frame. All frame-dependent kinematic variables (angles) are assumed to be generated in the CMS.

The lab frame corresponds to the system, where the target proton is at rest and the axis orientation is the following: $Z_{lab}$ -- along the beam, $Y_{lab}$ -- up, and $X_{lab}$ -- perpendicular to the $YZ$-plane.


For the scattered electron the task is rather straightforward. Its four-momentum can be calculated in the lab frame in the following way,

\begin{equation}
\begin{split}
  \nu &= \frac{W^2+Q^2-m_{p}^{2}}{2m_{p}}\\
  E_{e'} &= E_{beam}-\nu\\
 \theta_{e'} &= acos\left (1-\frac{Q^2}{2E_{beam}E_{e'}}\right )\\
P_{e'}^{4} = (E_{e'}sin \theta_{e'}cos \varphi_{e'}&,E_{e'}sin \theta_{e'}sin \varphi_{e'},E_{e'}cos \theta_{e'},E_{e'}),
\end{split}
\end{equation}\label{eq:el_in_lab}

where $\nu$ is the virtual photon energy in the lab frame, $m_{p}$ the target proton mass, $E_{beam}$ the incident electron beam energy,  which is among the input parameters, and $E_{e'}$ and $\theta_{e'}$ the scattered electron energy and polar angle, respectively. $W$, $Q^2$, and $\varphi_{e'}$ are the generated invariant mass of the final hadron system, the photon virtuality, and the azimuthal angle of the scattered electron, respectively.

For the final hadrons the task is significantly more complicated. To define their four-momenta in the lab frame~\footnote[2]{The calculation of hadron four-momenta in the lab frame is coded in subroutine {\em anti\_rot} in the file {\em anti\_rot.cxx}.} the following steps should be completed~\footnote[3]{In all derivations the energy is assumed to be the last component of the four-momentum and the four-momentum to be a row vector.}.

\begin{enumerate}

\item One should start with calculating the final hadron energies ($E_{1}$, $E_{2}$, $E_{3}$) and the magnitudes of their three-momenta ($P_{1}^{mag}$,$P_{2}^{mag}$, $P_{3}^{mag}$ ) in the CMS, since these variables do not depend on the axis orientation:

\begin{equation}
\begin{split}
E_{1} &= \frac{W^{2}+m_{1}^{2}-S_{23}}{2W},\;\;\;  P_{1}^{mag} = \sqrt{E_{1}^{2}-m_{1}^{2}},\\
E_{2} &= \frac{W^{2}+m_{2}^{2}-S_{13}}{2W},\;\;\;  P_{2}^{mag} = \sqrt{E_{2}^{2}-m_{2}^{2}},\\
E_{3} &= \frac{W^{2}+m_{3}^{2}-S_{12}}{2W},\;\;\;  P_{2}^{mag} = \sqrt{E_{3}^{2}-m_{3}^{2}},\\
\end{split}\label{eq:cms_prime_en_mom_mag}
\end{equation}

where $m_{1}$, $m_{2}$, $m_{3}$ are the masses of the first, second, and third hadrons, respectively, $S_{12}$, $S_{23}$, $S_{13}$ are the invariant masses squared of the corresponding hadron pairs.

\item To define the final hadron four-momenta completely one needs to calculate their spatial angles for some arbitrary coordinate system. For that purpose one chooses the coordinate system, which is denoted CMS$'$ and shown in Fig.~\ref{fig:ang_pl_gen}. The plane $A$ is the plane, where the initial proton, virtual photon, and the first hadron are located, $B$ is the plane, where all final hadrons are situated (the same planes $A$ and $B$ are shown in Fig.~\ref{fig:alpha_2nd_set}). This auxiliary coordinate system has the following axis orientation: $Z'$ -- along the first hadron, and $X'$ -- perpendicular to $Z'$ and situated in the plane $A$. It is convenient to choose this system, because in it the azimuthal angles of the remaining final hadrons ($\varphi_{2}$ and $\varphi_{3}$) are equal to (or differ by $\pi$ from) the generated angle $\alpha_{h}$, which is defined as the angle between the planes $A$ and $B$ (see Sect.~\ref{sect:cr_sect_exp} for detail).



 
In the CMS$'$ shown in Fig.~\ref{fig:ang_pl_gen} the polar and azimuthal angles of all final hadrons can be expressed by Eq.~\eqref{eq:cms_prime_spat_ang}.

\begin{equation}
\begin{aligned}
\theta_{1} &= 0,\;\;\; & \varphi_{1} = &~0 & \\
\theta_{2} &= acos\left (\frac{m_{1}^{2}+m_{2}^{2}+2E_{1}E_{2}-S_{12}}{2\sqrt{E_{1}^{2}-m_{1}^{2}}\sqrt{E_{2}^{2}-m_{2}^{2}}}\right ),\;\;\; & \varphi_{2} = &~\alpha_{h} & \\
\theta_{3} &= acos\left (\frac{m_{1}^{2}+m_{3}^{2}+2E_{1}E_{3}-S_{13}}{2\sqrt{E_{1}^{2}-m_{1}^{2}}\sqrt{E_{3}^{2}-m_{3}^{2}}}\right ),\;\;\; &
\varphi_{3} = & \begin{sqcases} 
\alpha_{h} + \pi,\;\;if\;\alpha_{h}<\pi \\ 
\alpha_{h} - \pi,\;\;if\;\alpha_{h}>\pi 
\end{sqcases} & \\
\end{aligned}\label{eq:cms_prime_spat_ang}
\end{equation}

\begin{figure}[htp]
\begin{center}
\includegraphics[width=12cm]{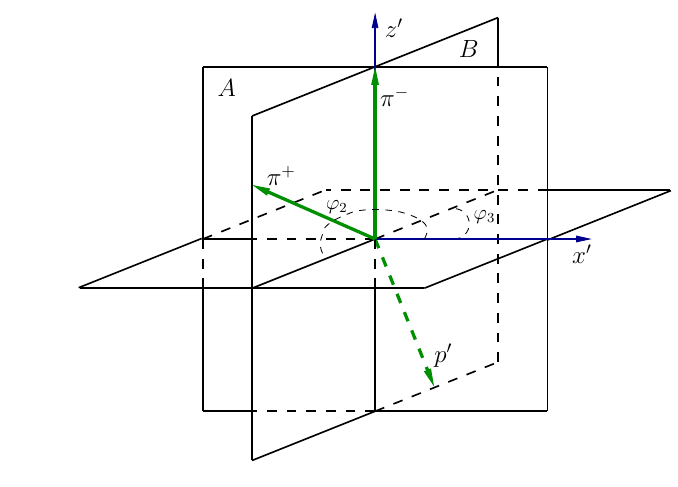}
\caption{\small Auxiliary frame CMS$'$ with the following axis orientation: $Z'$ -- along the first hadron and $X'$ -- perpendicular to $Z'$ and situated in the plane $A$. The plane $A$ is the plane, where the initial proton, virtual photon, and the first hadron are located, and $B$ is the plane, where all final hadrons are situated (the same planes $A$ and $B$ are shown in Fig.~\ref{fig:alpha_2nd_set}).  } \label{fig:ang_pl_gen}
\end{center}
\end{figure}

The spatial angles defined by Eq.~\eqref{eq:cms_prime_spat_ang} together with the energies and momentum magnitudes defined by Eq.~\eqref{eq:cms_prime_en_mom_mag} define completely the four-momenta of final hadrons in the auxiliary frame CMS$'$.

\item Once the four-momenta of the final hadrons are defined in the auxiliary coordinate system CMS$'$ shown in Fig.~\ref{fig:ang_pl_gen}, one can transform them into the coordinate system with the usual CMS-axis-orientation, where $Z_{cms}$ -- along the virtual photon, $X_{cms}$ -- in the electron scattering plane $(e,~e')$, and $Y_{cms}$ -- perpendicular to the $XZ$-plane. For that purpose two subsequent rotations should be made.

First, the $Z'$-axis of the auxiliary system in Fig.~\ref{fig:ang_pl_gen} is rotated with the generated angle $\theta_{h}$ in the plane $A$ to set  $Z'$-axis along the virtual photon. This rotation transforms the four-momentum as $P' = P\cdot R_{\theta}$, with 

\begin{equation}
R_{\theta}=\begin{pmatrix}
cos\theta_{h} &0  &-sin\theta_{h}  &0 \\ 
 0& 1 & 0 &0 \\ 
 sin\theta_{h} &0  &cos\theta_{h}  & 0\\ 
0 &0  & 0 &1 
\end{pmatrix}.
\end{equation}

After this rotation the $X'$-axis is still in the plane $A$.

Then one should rotate the $X'$-axis with the generated angle $\varphi_{h}$ in the $X'Y'$-plane (around the $Z'$-axis) to bring the $X'$-axis into the $(e,~e')$-plane. This rotation transforms the four-momentum as $P'' = P'\cdot R_{\varphi}$, with 

\begin{equation}
R_{\varphi} = \begin{pmatrix}
 cos\varphi_{h}& sin\varphi_{h} & 0 &0 \\ 
 -sin\varphi_{h}& cos\varphi_{h} &  0& 0\\ 
0 & 0 & 1 &0 \\ 
 0&  0&  0&1 
\end{pmatrix}.
\end{equation}

After that the final hadron four-momenta are defined in the CMS with the usual axis orientation.

\item Now one can perform the boost from the CMS to the lab$'$ frame. The prime means that although the proton is at rest in this system, the axis orientation is still like in the CMS. The boost is given by $P''' = P''\cdot R_{boost}$, with 
\begin{equation}
R_{boost} = \begin{pmatrix}
1 &0  &0  &0 \\ 
0 &1  &0  &0 \\ 
 0&  0& \gamma  &\gamma \beta  \\ 
 0&  0& \gamma \beta  & \gamma 
\end{pmatrix}, \, \, \, \beta =\frac{\sqrt{\nu^2+Q^{2}}}{\nu+m_{p}}, \, \, \,  \textrm{and} \,\,\,   \gamma =\frac{1}{\sqrt{1-\beta ^{2}}},
\end{equation}
where $\nu$ is the energy of the virtual photon in the lab frame, which is defined in Eq.~\eqref{eq:el_in_lab}, and $\beta$ the magnitude and $z$-component of the three-vector $\overrightarrow{\beta}=(0,0,\beta)$.~\footnote[4]{
Note: if you use ROOT function .Boost the positive sign should be assigned to the $z$-component of $\beta$-vector, i.e. .Boost(0,0,$\beta$).}

After the boost the four-momenta of the final hadrons are defined in the lab$'$ frame, which has the CMS-like axis orientation.

\item The only thing left is to rotate the coordinate axes into the usual lab frame axis orientation.

\begin{figure}[htp]
\begin{center}
\includegraphics[width=6cm]{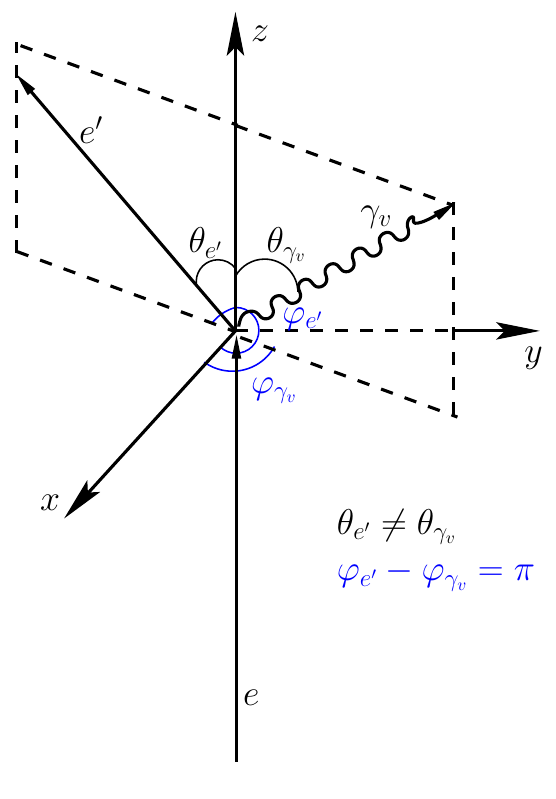}
\caption{\small Virtual photon and scattered electron angles $\theta$ and $\varphi$ in the lab frame.} \label{fig:el_angles}
\end{center}
\end{figure}

For that purpose one firstly should define in the lab frame the vectors $\overrightarrow{n}_{e'}^{lab}$ and $\overrightarrow{n}_{\gamma}^{lab}$, which are the unit vectors along the scattered electron and virtual photon, respectively. The spatial angles of the scattered electron and virtual photon are shown in Fig.~\ref{fig:el_angles}. 

\begin{equation}
\overrightarrow{n}_{e'}^{lab} = sin\theta_{e'}cos\varphi_{e'}\overrightarrow{i} + sin\theta_{e'}sin\varphi_{e'}\overrightarrow{j} + cos\theta_{e'}\overrightarrow{k},
\end{equation}

where $\theta_{e'}$ is defined in Eq.~\eqref{eq:el_in_lab} and $\varphi_{e'}$ is the generated randomly value (see Sect.~\ref{sect:gen_with_weights}), $\overrightarrow{i}$, $\overrightarrow{j}$, $\overrightarrow{k}$ are the unit vectors along the $x$, $y$, $z$ axes of the lab system, respectively.

\begin{equation}
\begin{split}
\theta_{\gamma} &= acos \left ( \frac{Q^2+2E_{beam}\nu}{2E_{beam}\sqrt{Q^2+\nu^2}} \right )\\
\varphi_{\gamma} &= \begin{sqcases} 
\varphi_{e'} + \pi,\;\;if\;\varphi_{e'}<\pi \\ 
\varphi_{e'} - \pi,\;\;if\;\varphi_{e'}>\pi 
\end{sqcases}\\
\overrightarrow{n}_{\gamma}^{lab} &= sin\theta_{\gamma}cos\varphi_{\gamma}\overrightarrow{i} + sin\theta_{\gamma}sin\varphi_{\gamma}\overrightarrow{j} + cos\theta_{\gamma}\overrightarrow{k}.
\end{split}
\end{equation}

The idea of the coordinate system transformation is that the unit axis-vectors $\overrightarrow{u}$, $\overrightarrow{v}$, $\overrightarrow{w}$ of the system lab$'$ should be written in the lab system:

\begin{equation}
\begin{split}
\overrightarrow{w} &= \overrightarrow{n}_{\gamma}^{lab} = sin\theta_{\gamma}cos\varphi_{\gamma}\overrightarrow{i}+sin\theta_{\gamma}sin\varphi_{\gamma}\overrightarrow{j}+cos\theta_{\gamma}\overrightarrow{k}\\
\overrightarrow{v} &= [\overrightarrow{n}_{\gamma}^{lab}\times \overrightarrow{n}_{e'}^{lab}] = \begin{vmatrix}
\overrightarrow{i} & \overrightarrow{j} & \overrightarrow{k}\\ 
sin\theta_{\gamma}cos\varphi_{\gamma} & sin\theta_{\gamma}sin\varphi_{\gamma} & cos\theta_{\gamma}\\ 
sin\theta_{e'}cos\varphi_{e'} &sin\theta_{e'}sin\varphi_{e'}  &cos\theta_{e'} 
\end{vmatrix}\\
\overrightarrow{u} &= [\overrightarrow{v}\times \overrightarrow{w}].
\end{split}
\end{equation}

\vspace{5em}
The four-momenta are transformed as $P'''' = P'''\cdot R_{lab}$~\footnote[5]{Using embedded ROOT functions, this transformation can be coded via Euler angles in the following way. \\ 
TRotation rot; \\
rot.SetXEulerAngles$( atan2\left (u_{z},v_{z} \right ),acos\left ( w_{z} \right ),atan2 \left (  w_{x},-w_{y} \right ) )$; \\
P.Transform(rot);
}, with
 
\begin{equation}
R_{lab} = \begin{pmatrix}
 u_{x}& u_{y} & u_{z} &0 \\ 
 v_{x}& v_{y} & v_{z} & 0\\ 
 w_{x}& w_{y} &  w_{z} &0 \\ 
 0&  0&  0&1 
\end{pmatrix}.
\end{equation}
\end{enumerate}

After all steps are completed the four-momenta of the final hadrons are written in the lab frame.

\chapter{Obtaining the weights}
\label{sect:data}
\section{Data used to obtain the weights}
\mbox{}\vspace{-\baselineskip}

The following sets of data were included into the EG.

\begin{enumerate}
\item Electroproduction data:
\begin{enumerate}
\item CLAS data at $E_{beam} = 2.445$ GeV,  $E_{beam} = 4$ GeV \cite{Ripani:2002ss}, \cite{Mokeev:2015lda};
\item CLAS data at $E_{beam} = 1.515$ GeV \cite{Fedotov:2008aa}, \cite{Mokeev:2008iw}, \cite{Mokeev:2012vsa}.
\end{enumerate}
\item Photoproduction data:
\begin{enumerate}
\item CLAS g11a data \cite{Golovach:note};
\item SAPHIR \& ABBHM data on integral cross sections \cite{Wu:2005wf}, \cite{ABBHHM:1968aa}.  
\end{enumerate}
\end{enumerate}

Single-differential electroproduction cross sections 1a~\cite{Ripani:2002ss} and 1b~\cite{Fedotov:2008aa} were fit with the JM model~\cite{Mokeev:2015lda},\cite{Mokeev:2008iw},\cite{Mokeev:2012vsa}, therefore all structure functions in Eq.~\eqref{eq:str_fun_decomp} obtained in this fit in a five-dimensional sense are used in TWOPEG.

Single-differential photoproduction cross sections 2a~\cite{Golovach:note} were also fit with the JM model that leads to the fact that at photon point $\sigma_{T}$ is also available from the fit in a five-dimensional sense. This is not the case for data-set 2b~\cite{Wu:2005wf}, for which only integral values of experimental cross section are available for different $W$ bins.

\begin{figure}[htp]
\begin{center}
\includegraphics[width=16cm]{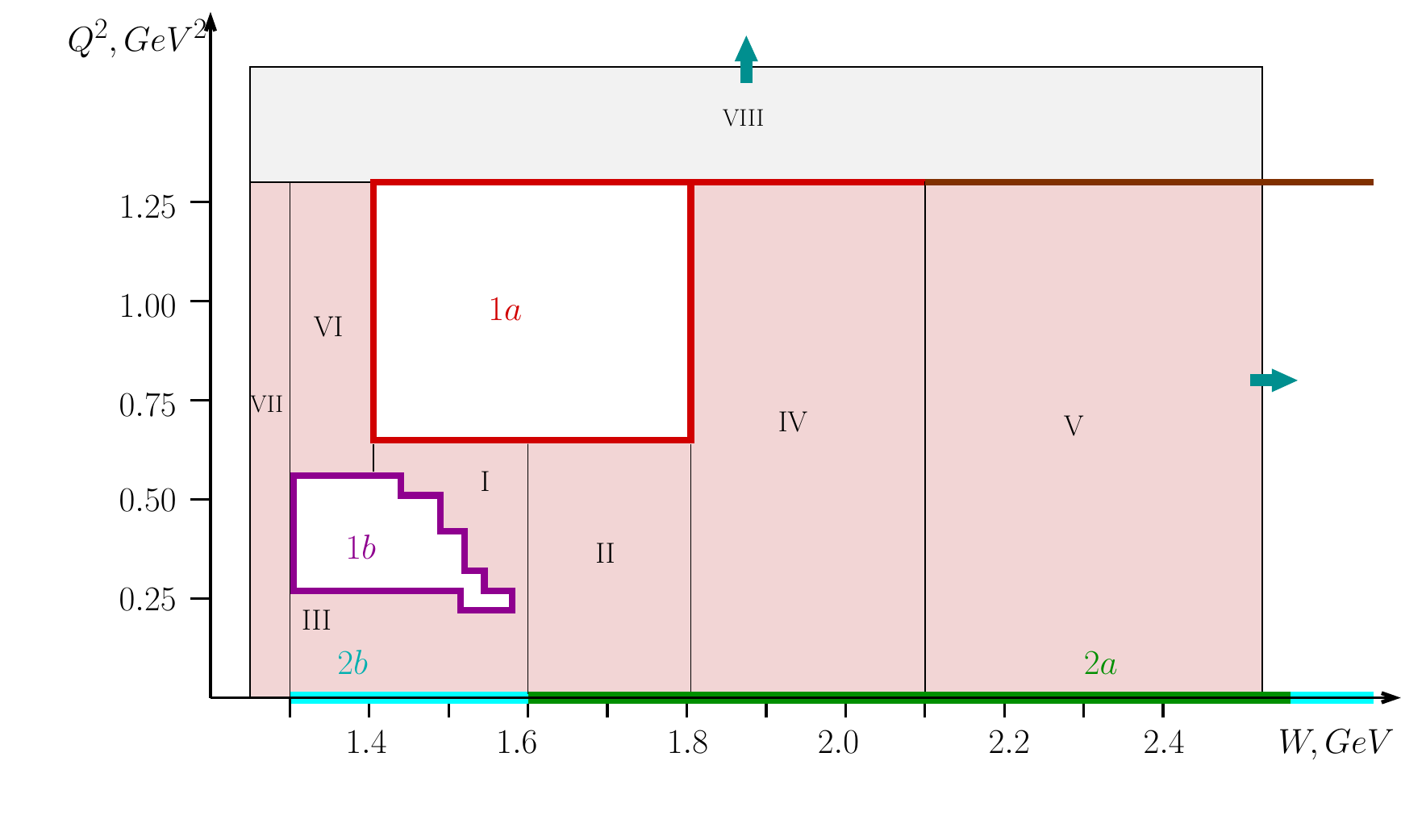}
\caption{\small $W-Q^2$ coverage of TWOPEG. The areas within red and lilac boundaries correspond to the electroproduction data 1a and 1b, respectively. The green and cyan lines along the horizontal $W$-axis correspond to the photoproduction data 2a and 2b, respectively. The brown line at $Q^2 = 1.3$ GeV$^2$ corresponds to the area for which the cross section was roughly estimated  from the JM model in a five-dimensional sense without relying on experimental data. The blue arrows indicate the extension to the higher $W$ and $Q^2$.} \label{fig:gen_cover}
\end{center}
\end{figure}

Figure~\ref{fig:gen_cover} shows which $W-Q^2$ areas are covered by these data-sets. The areas within red and lilac boundaries correspond to the electroproduction data 1a and 1b, respectively. The green and cyan lines along the horizontal $W$-axis correspond to the photoproduction data 2a and 2b, respectively. As it is seen in Fig.~\ref{fig:gen_cover} information about double pion production cross section exists only in very limited regions, while the major part of the $W-Q^2$ plane lacks information. In these areas, where information about double pion production cross section does not exist special interpolation and extrapolation procedures have been developed and applied in order to obtain reasonable cross section estimates (see Sect.~\ref{sect:cr_sect_extr_intr}). The brown line at $Q^2 = 1.3$ GeV$^2$ corresponds to the area for which the cross section was roughly estimated  from the JM model in a five-dimensional sense without relying on experimental data. It was done in order to extend the $W-Q^2$ coverage of TWOPEG further to the region of higher $W$-values, where no experimental data exists up to now.

\section{Calculation of weights}
\label{sect:cr_sect_extr_intr}


As it was mentioned in the previous section the single-fold differential cross sections from the data-sets 1a, 1b, and 2a were fit with the JM model, and as a result all structure functions from Eq.~\eqref{eq:str_fun_decomp} are available in a five-dimensional sense (from the photoproduction data-set 2a only $\sigma_{T}$ exists). Table~\ref{tab:mod_grid} shows the kinematical grid~\footnote[1]{Note that although the model cross section is differential in $(-cos \theta_{h})$, a grid that is equidistant in $\theta_{h}$ is used.}, in which the model structure functions were obtained.

\begin{table}[h]
\begin{center}
  \renewcommand{\arraystretch}{1.2}
\begin{tabular}{|c|c|c|c|c|c|c|c|}

\hline
Data & $Q^2$ values, & \# of $W$  & \# of $S_{12}$ & \# of $S_{23}$ & \# of $\theta_{h}$ & \# of $\alpha_{h}$  & \# of $\varphi_{h}$ \\
 set &  GeV$^2$ &  points &  points& points &  points &  points &  points\\
\hline
   &  0.65 & 17 &    &    &   &   & \\
1a &  0.95 & 17 & 12 & 12 & 6 & 6 & 1\\
   &  1.30 & 28 &    &    &   &   & \\
\hline
   &  0.225 & 4  &   &    &   &   &  \\
   &  0.275 & 11 &   &    &   &   &  \\
   &  0.325 & 10 &   &    &   &   &  \\
1b &  0.425 &  9 & 10&10  & 8 & 8 & 1\\
   &  0.475 &  8 &   &    &   &   &  \\
   &  0.525 &  8 &   &    &   &   &  \\
   &  0.575 &  6 &   &    &   &   &  \\
\hline
2a &  0.   &  30& 16&16  & 14&14 & 1\\
\hline
\end{tabular}
\caption{\small Kinematical grid, in which the model structure functions are obtained. \label{tab:mod_grid}}
\end{center}
\end{table}

To  calculate the weight $f_{cr~sect}$ (see Eq.~\eqref{eq:weight_cr_sect}), means to estimate the value of the five-differential hadronic double pion cross section in the given corresponding 7-dimensional kinematical point ($W$,~$Q^2$,~$S_{12}$, $S_{23}$,~$\theta_{h}$,~$\varphi_{h}$,~$\alpha_{h}$). It in turn means that all structure functions from Eq.~\eqref{eq:str_fun_decomp} need to be estimated in that point.  The procedure of interpolation and extrapolation of the model structure functions into the desired point is described next.

First, for each ($W$,~$Q^2$) point, where model structure functions are available, the four-dimensional linear interpolation is done between the points of the ($S_{12}$,~$S_{23}$,~$\theta_{h}$,~$\alpha_{h}$) grid. Then for the electroproduction data a two-dimensional linear interpolation is done between the points of the ($W$,~$Q^2$) grid, while for the photoproduction data a one-dimensional linear $W$-interpolation is done. As a result one can get the values of the structure functions for any arbitrary point located in the inner areas within the red and lilac boundaries and along the green line in Fig.~\ref{fig:gen_cover}. 

Region I. The two-dimensional linear ($W$,~$Q^2$) interpolation is also performed between the regions 1a and 1b for $1.4 < W < 1.6$ GeV. 

In order to estimate the values of the structure functions in other regions (colored by pink in Fig.~\ref{fig:gen_cover} and numbered by roman numerals) the special procedures should be applied.

Region II. Let's consider the region $1.6 < W < 1.825$ GeV and $0 < Q^2 < 0.65$ GeV$^2$. For $Q^2 = 0.65$~GeV$^2$ the full set of five-dimensional structure functions are available from the model analysis of data-set 1a. For the photon point only $\sigma_{T}$ is available from the model analysis of data-set 2a. In order to estimate the value of all structure functions in any point of this region a procedure, which contains the following steps was developed.

\begin{itemize}

\item The integral values of model structure function $\sigma_T$ in three $Q^2$ points (0, 0.65, and 0.95~GeV$^2$) were fit with the following function for each $W$ point individually:
\begin{equation}
F_{T}(Q^2) = a_{0}\left (Q^2+a_1 \right )^{a_2}+a_{3},
\label{eq:q2_dep_fit_sig_t}
\end{equation}

where $a_{0}$, $a_{1}$, $a_{2}$, and $a_{3}$ are free fit parameters.

The result of this fit for $W = 1.7125$ GeV is shown in Fig.~\ref{fig:sig_t_l_fit} (left side). The black points are the integrated values of model structure function $\sigma_{T}$ and the red curve is the fit function $F_{T}(Q^2)$.

Then the five-differential structure function $\sigma_{T}$ can be scaled between the $Q^2$ boundaries to any arbitrary $Q^2$ point along the obtained $Q^2$ dependence. But since $\sigma_{T}$ exists in a five-dimensional sense at both boundaries ($Q^2 = 0.65$~GeV$^2$ and $Q^2 = 0$~GeV$^2$), it can be scaled from both sides. To avoid loss of information about the evolution of the differential cross section shape with $Q^2$, it was decided to mix these two scaled structure functions at each $Q^2$ point. Thus the differential structure function in arbitrary point $0 < Q^2 < 0.65$~GeV$^2$ is given by

\begin{equation}
\frac{d^5\sigma_{T}}{d^5\tau}\left ( Q^2 \right ) = \frac{Q^2}{0.65}\frac{d^5\sigma_{T}}{d^5\tau}\left ( 0.65 \right )\frac{F_{T}(Q^2)}{F_{T}(0.65)}+\frac{0.65-Q^2}{0.65}\frac{d^5\sigma_{T}}{d^5\tau}\left ( 0 \right )\frac{F_{T}(Q^2)}{F_{T}(0)}.
\label{eq:q2_dep_fit_sig_t}
\end{equation}

The first term corresponds to the value of $\sigma_{T}$, which is scaled from the $Q^2=0.65$~GeV$^2$ edge, while the second one is scaled from the $Q^2 = 0$~GeV$^2$ edge. The first fraction in each term corresponds to the mixing coefficient, which forces the term to completely dominate at its own edge and to vanish at the opposite edge. The last fraction in each term corresponds to the integral scaling. All numerical values in Eq.~\eqref{eq:q2_dep_fit_sig_t} are given in GeV$^2$.

\begin{figure}[htp]
\begin{center}
\includegraphics[height=0.32\textwidth]{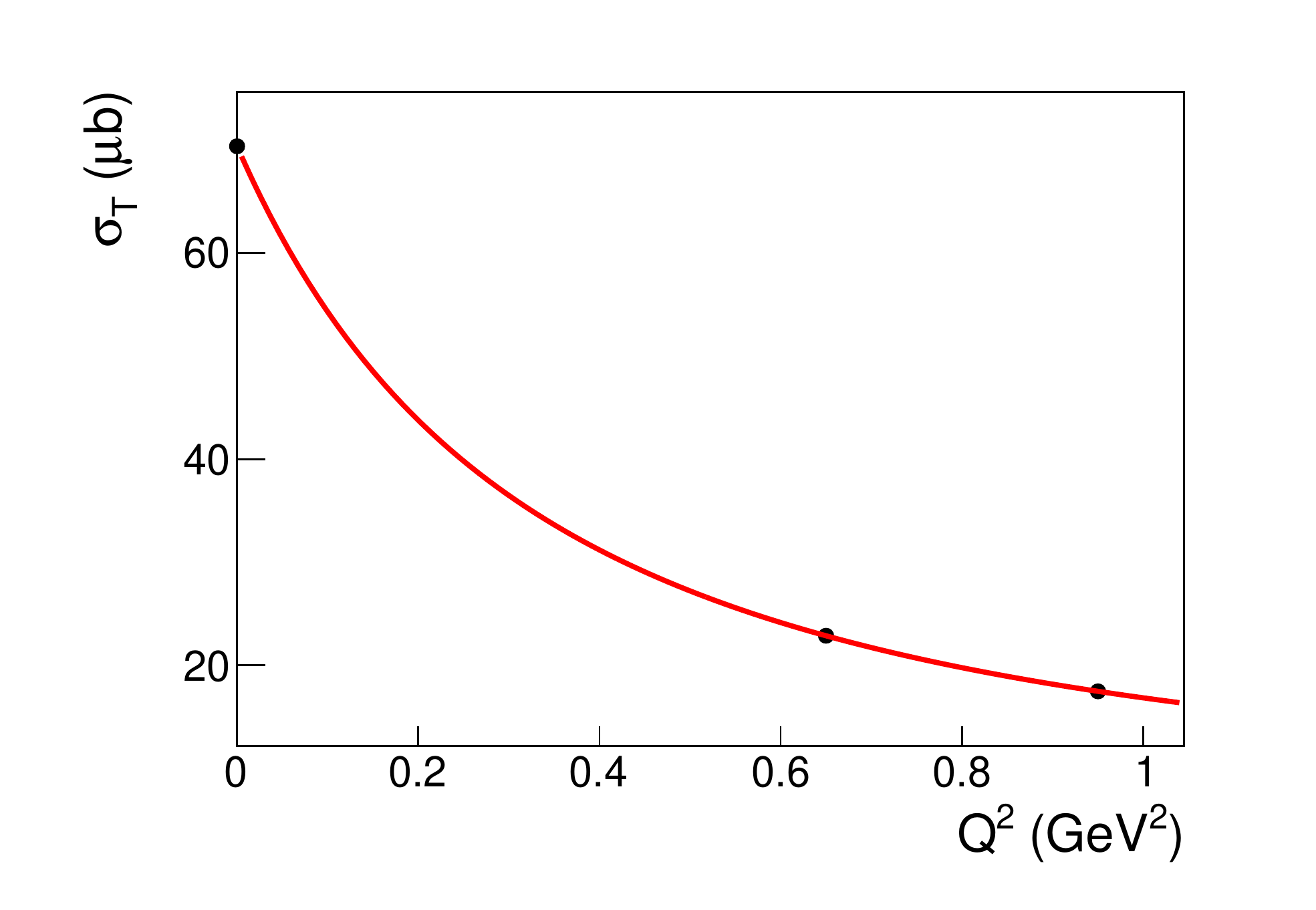}
\includegraphics[height=0.32\textwidth]{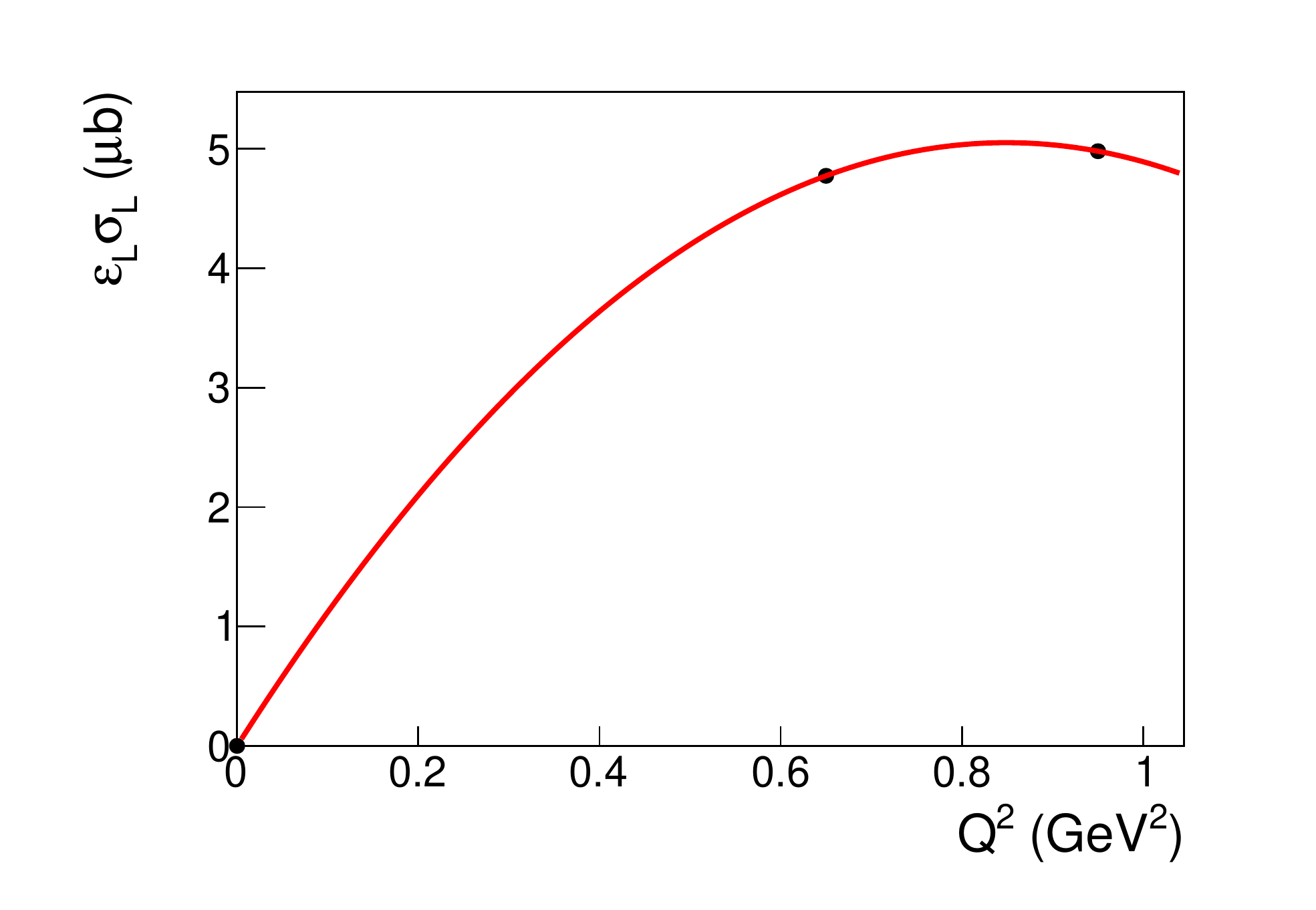}
\caption{\small Left side: fit of the integrated values of model structure function $\sigma_{T}$ (black points) with the function given by Eq.~\eqref{eq:q2_dep_fit_sig_t} for the $W = 1.7125$ GeV. The red curve corresponds to the function $F_{T}(Q^2)$, which is the fit result. Right side: fit of $\varepsilon_{L}\sigma_L$ with a second order polynom for $W = 1.7125$ GeV. The values of $\varepsilon_{L}$ are calculated for a beam energy of 2.445~GeV, with which the experiment 1a ran. The red curve corresponds to the function $F_{L}(Q^2)$, which is the fit result. } \label{fig:sig_t_l_fit}
\end{center}
\end{figure}

\item For the structure function $\sigma_{L}$ the situation is more complicated, since no information about it exists at the photon point and for this reason one can neither perform the fit nor the cross section mixing as described in the previous step. Hence the procedure should be modified. The modification is made based on the fact that although the information about the $\sigma_{L}$ behavior close to the photon point is unknown, the combination $\varepsilon_{L}\sigma_{L}$ must definitely vanish at $Q^2 = 0$~GeV$^2$. Therefore, the values of the combination $\varepsilon_{L}\sigma_L$ in the three $Q^2$ points (0, 0.65, and 0.95~GeV$^2$) were fit with a second order polynom. The result of this fit for $W = 1.7125$ GeV is shown in Fig.~\ref{fig:sig_t_l_fit} (right side). The black points are the values of $\varepsilon_{L}\sigma_L$, where $\sigma_{L}$ is the integrated model structure function and $\varepsilon_{L}$ is defined in the context of Eq.~\eqref{eq:str_fun_decomp}. The red curve is the resulting fit function $F_{L}(Q^2)$.

Figure~\ref{fig:sig_gen} shows the event distributions of TWOPEG, which are obtained based on this approach. Left plot shows the $Q^2$ dependence of the integral  quantities $\sigma_{T}$ (blue curve), $\varepsilon_{L}\sigma_{L}$ (green curve) and $\sigma_{T}+\varepsilon_{L}\sigma_{L}$ (red curve) in comparison with the model points. As it is seen in this plot the longitudinal term $\varepsilon_{L}\sigma_{L}$ gives rather small contribution to the full cross section. Right plot shows the estimated behavior of the structure function $\sigma_{L}$. 

\begin{figure}[htp]
\begin{center}
\includegraphics[height=0.35\textwidth]{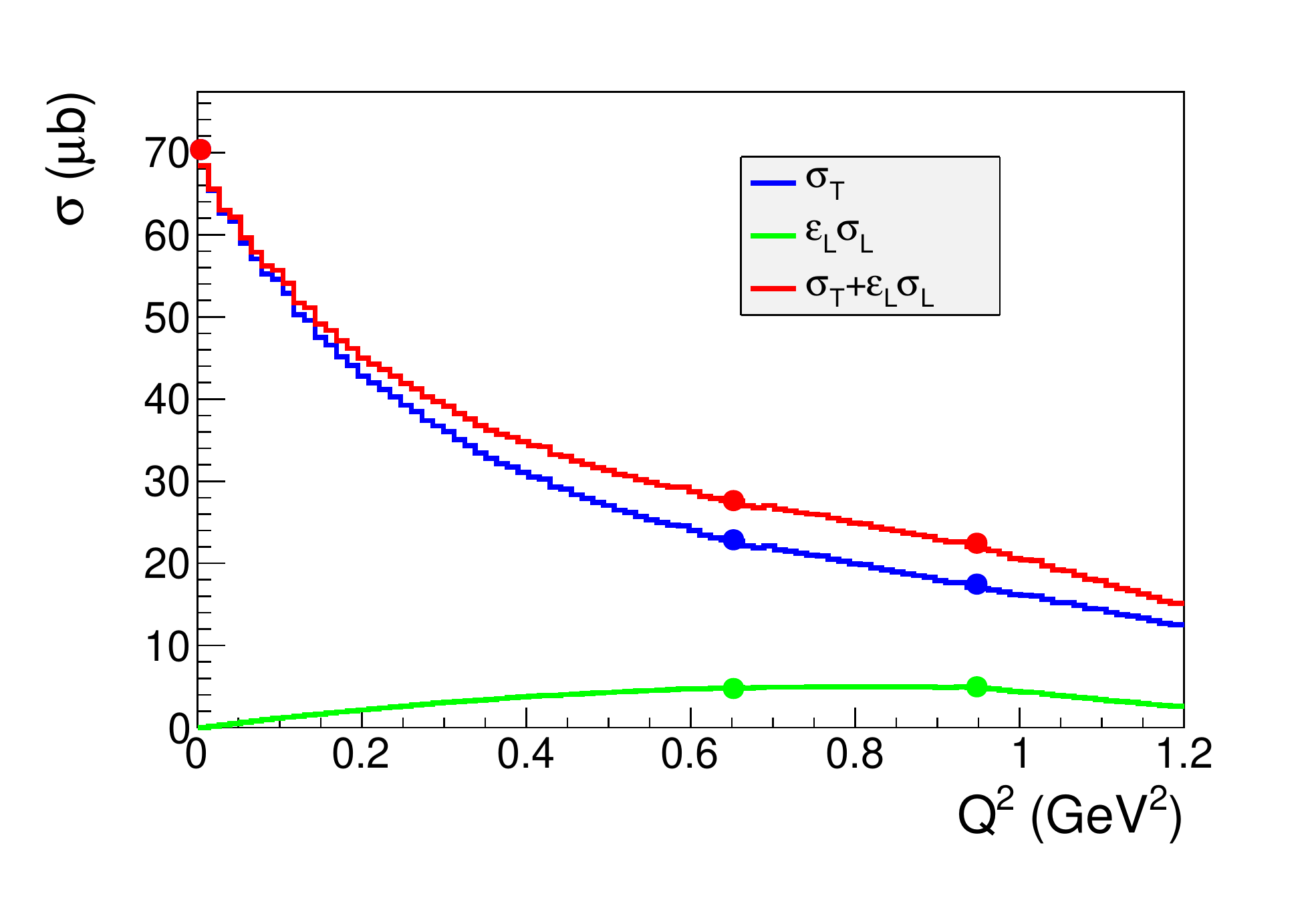}
\includegraphics[height=0.35\textwidth]{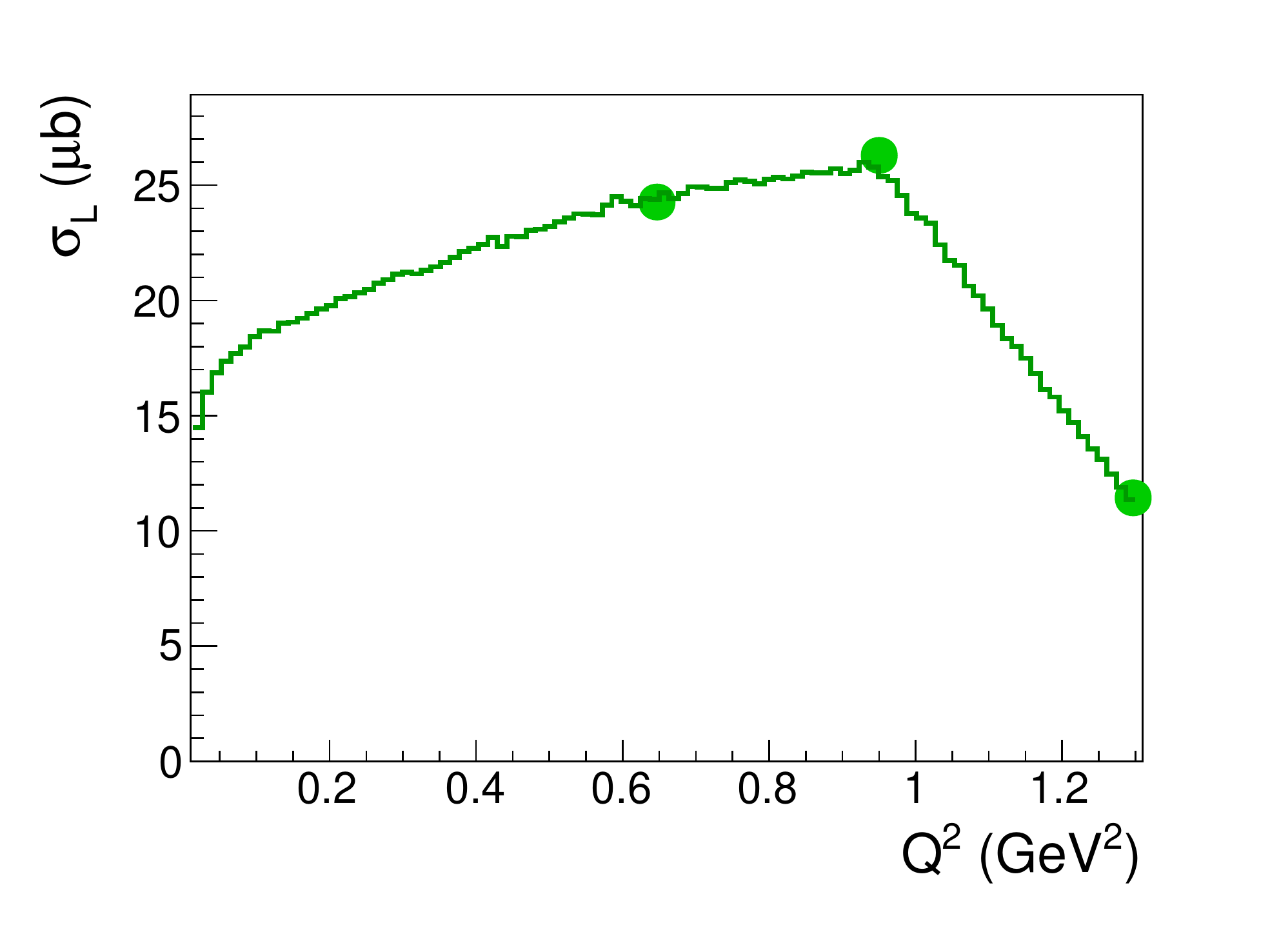}
\caption{\small $Q^2$ dependence of different contributors to the full integral cross sections. Comparison between the event distributions of TWOPEG (curves) with the JM model values (points). Left side: integral $\sigma_{T}$ (blue), $\varepsilon_{L}\sigma_{L}$ (green), and $\sigma_{T}+\varepsilon_{L}\sigma_{L}$ (red) as a function of $Q^2$. The values of $\varepsilon_{L}$ are calculated for a beam energy of 2.445~GeV, with which the experiment 1a ran. Right side: the structure function $\sigma_{L}$ as a function of $Q^2$. For both plots curves are obtained with TWOPEG for the $W$-bin [1.7, 1.725]~GeV, while the points are for the JM model at $W=1.7125$~GeV.  } \label{fig:sig_gen}
\end{center}
\end{figure}

Since $\sigma_{L}$ exists in a five-dimensional sense at $Q^2 = 0.65$~GeV$^2$ the differential structure function $\sigma_{L}$ in any point at $0 < Q^2<0.65$~GeV$^2$ can be obtained by

\begin{equation}
\frac{d^5\sigma_{L}}{d^5\tau}\left ( Q^2 \right ) = \frac{d^5\sigma_{L}}{d^5\tau}\left ( 0.65\;{\rm GeV^2} \right )\frac{F_{L}(Q^2)}{F_{L}(0.65\;{\rm GeV^2})}\frac{\varepsilon_{L}(2.445\;{\rm GeV},0.65\;{\rm GeV^2})}{\varepsilon_{L}(2.445\;{\rm GeV},Q^2)},
\label{eq:q2_dep_fit_sig_l}
\end{equation}

where the value of $\varepsilon_{L}$ is calculated here for $E_{beam} = 2.445$~GeV in order to satisfy the conditions, under which the experiment 1a was made.

\item The terms in Eq.~\eqref{eq:str_fun_decomp} that correspond to other structure functions are also forced by the coefficients in front of them to vanish at $Q^2 = 0$~GeV$^2$. Hence these structure functions exist in five-dimensional sense only at the $Q^2 = 0.65$~GeV$^2$ edge. Their contribution into the full cross section is an order of a magnitude lower than the contribution from the first (transverse) and second (longitudinal) terms, thus there is no need to develop complex procedure for the purpose of their extrapolation to the region $Q^2 < 0.65$~GeV$^2$. For this purpose the following approximate $Q^2$ dependence is used:

 \begin{equation}
F_{approx}(Q^2) = \frac{1}{\left (1+\frac{Q^2}{0.7~GeV^2}\right )^{2}},
\label{eq:q2_dep}
\end{equation}

The differential structure functions in any point at $0 < Q^2<0.65$~GeV$^2$ can be obtained in this way:

 \begin{equation}
\frac{d^5\sigma_{X}}{d^5\tau}\left ( Q^2 \right ) = \frac{d^5\sigma_{X}}{d^5\tau}\left ( 0.65\;{\rm GeV^2} \right )\frac{F_{approx}(Q^2)}{F_{approx}(0.65\;{\rm GeV^2})},
\label{eq:sig_x_acale}
\end{equation}

where $\sigma_X$ corresponds to $\sigma_{TT}$, $\widetilde{\sigma}_{TT}$, $\sigma_{TL}$, and $\widetilde{\sigma}_{TL}$.

\item Last step of the procedure is to perform linear one-dimensional interpolation of all structure functions into the given $W$ point.

\end{itemize}

After that in any arbitrary point in the region of $1.6 < W < 1.825$ GeV and $0<Q^2<0.65$ GeV$^2$ all structure functions are estimated in a multi-dimensional sense.

Region III. In the region of $1.3 < W < 1.6$ GeV and $0 < Q^2 < 0.275$ GeV$^2$ the same procedure as for region II is applied, with the exception of $\sigma_{T}$ mixing, which can not be done, since structure function $\sigma_{T}$ is available in a five-dimensional sense only at $Q^2 = 0.275$~GeV$^2$, and at $Q^2 = 0$~GeV$^2$ just integral values of $\sigma_{T}$ are known from the data-set 2b (cyan line in Fig.~\ref{fig:gen_cover}). This leads to the fact that the shapes of all differential structure functions for $Q^2 < 0.275$~GeV$^2$ remain the same as at  $Q^2 = 0.275$~GeV$^2$, although their absolute normalization changes properly due to the integral scaling.

Region IV. Now let's proceed to the region of $1.825 < W < 2.1$ GeV and $0<Q^2<1.3$ GeV$^2$. For $Q^2 = 1.3$~GeV$^2$ the full set of five-dimensional structure functions are available from the model analysis of data-set 1a. For the photon point only $\sigma_{T}$ is available from the model analysis of the data-set 2a. Therefore, a procedure similar to that described above for the region II can be applied here. The only difference is that the $Q^2$ dependence of all structure functions was not obtained by the fit of the corresponding integral values, but was taken to be the same as the one at $W = 1.8125$~GeV. This was done for the following reasons. A fit of the integral $\sigma_{T}$ structure function is not expected to be reliable, since for $\sigma_{T}$ model prediction exists only for two $Q^2$ points (at $Q^2 = 1.3$~GeV$^2$ and $Q^2 = 0$~GeV$^2$), which are very far from each other. For the longitudinal part of the cross section no fit can be done, since the model prediction exits only at a single point, $Q^2 = 1.3$~GeV$^2$. 

Region V. The same procedure (IV) is also used for the region of $W > 2.1$ GeV and $0<Q^2<1.3$ GeV$^2$ with the only distinction that for $W > 2.1$~GeV and $Q^2 = 1.3$~GeV$^2$ the five-dimensional structure functions were roughly estimated by the JM model without relying on experimental data, since no data exists there (the brown line in Fig.~\ref{fig:gen_cover}). The same was done also for $W > 2.5$~GeV and $Q^2 = 0$~GeV$^2$, but here the information about integral photoproduction cross sections (data-set 2b shown with the cyan line in Fig.~\ref{fig:gen_cover}) was used. Thereby TWOPEG works up to $W = 4.5$~GeV, but the higher the value of $W$ is, the less reliable the cross section shape becomes.

Region VI. For the region of $1.3 < W < 1.4$~GeV and $0.575 < Q^2 < 1.3$~GeV$^2$ the five-differential structure functions, which exist at $Q^2 = 0.575$~GeV$^2$ from the model analysis of data-set 1b, were scaled to the region $Q^2 > 0.575$~GeV$^2$ with the $Q^2$ dependence that is taken to be the same as at $W = 1.4125$~GeV. This dependence is linear between the $Q^2$ points 0.575, 0.65, 0.95, and 1.3~GeV$^2$. Thus the shape of all differential structure functions remains the same in this region, and only integral scaling takes place.

Region VII. In order to extend the EG coverage closer to the threshold the following was done in the region of $1.2375 < W < 1.3125$~GeV and $0.005 < Q^2 < 1.3$~GeV$^2$. Three extra $W$ points at 1.2375, 1.2625, and 1.2875 GeV were added. The structure functions at these points are taken to be the same as at the point $W = 1.3125$~GeV, except for two modifications: 1) invariant mass distributions are forced to shrink as $W$ decreases and 2) integral scaling is applied in order to force the integral cross section vanish at the threshold. The $Q^2$ dependence remains the same as at $W=1.3125$~GeV. A linear $W$ interpolation is done between the additional $W$ points.

Region VIII. In regard to the extension of the EG coverage to the region with $Q^2 > 1.3$~GeV$^2$ the simple approximate $Q^2$ dependence given by Eq.~\eqref{eq:q2_dep} is used. The structure functions in this case are given by

\begin{equation}
\frac{d^5\sigma_{X}}{d^5\tau}\left ( Q^2  > 1.3~{\rm GeV^2} \right ) = \frac{d^5\sigma_{X}}{d^5\tau}\left ( 1.3\;{\rm GeV^2} \right )\frac{F_{approx}(Q^2)}{F_{approx}(1.3\;{\rm GeV^2})},
\label{eq:sig_x_acale_gt_1_3}
\end{equation}

where $\sigma_X$ corresponds to $\sigma_{T}$, $\sigma_{L}$, $\sigma_{TT}$, $\widetilde{\sigma}_{TT}$, $\sigma_{TL}$ and $\widetilde{\sigma}_{TL}$.

The shape of all differential structure functions for $Q^2 > 1.3$~GeV$^2$ remains the same as at  $Q^2 = 1.3$~GeV$^2$, but their absolute values decrease according to the integral scaling given by Eq.~\eqref{eq:sig_x_acale_gt_1_3}.

After all these steps adapted to the different $W-Q^2$ regions shown in Fig.~\ref{fig:gen_cover} are applied, all structure functions can be obtained at any arbitrary point for $W$ in [1.2375, 4.5375]~GeV and $Q^2 > 0.005$~GeV$^2$. The resulting structure functions can be combined into the full five-differential hadronic cross section according to Eq.~\eqref{eq:str_fun_decomp} for any particular beam energy given as an input parameter. To obtain the weight factor for each generated event, this cross section must be multiplied by the virtual photon flux given by Eq.~\eqref{flux} in order to obtain the proper electron scattering cross section, which leads to

\begin{equation}
f_{cr~sect} = 2\pi^3(S_{12}^{max}-S_{12}^{min})(S_{23}^{max}-S_{23}^{min})\Gamma_{v}(W,Q^2)\frac{d^{5}\sigma_{v} }{d^{5}\tau},
\label{eq:weight_cr_sect}
\end{equation}
where $\frac{d^{5}\sigma_{v} }{d^{5}\tau}$ is the full hadronic cross section given by Eq.~\eqref{eq:str_fun_decomp} and $\Gamma_{v}(W,Q^2)$ the virtual photon flux given by Eq.~\eqref{flux}. Other factors are introduced in order to make the generated events consistent with the experimental ones and to force them to give the integral cross section value upon a weighted summation of events in $W$, $Q^2$ bin. 
The $W$ dependent factors $(S_{12}^{max}-S_{12}^{min})$ and $(S_{23}^{max}-S_{23}^{min})$ are especially essential for the correct weight propagation. $S_{12}^{min},~S_{12}^{max},~S_{23}^{min},~S_{23}^{max}$ are the minimal and maximal values of the corresponded invariant mass squared at the given $W$ point.   With these factors  one can get the proper $W$ dependence of the integrated cross section automatically by the weighted sum of events in the particular $W$ bins.

It also needs to be mentioned that the multiplication by the virtual photon flux $\Gamma_{v}$ in Eq.~\eqref{eq:weight_cr_sect} corresponds to the mode, where the input flag $F_{flux} = 1$. In this case TWOPEG simulates events that are collected during the electron scattering experiments. This mode must be used if the EG is used for the modeling of experiment, for example efficiency evaluation.

But TWOPEG can also work in the mode $F_{flux} = 0$, in which the weight $f_{cr~sect}$ is not multiplied by the virtual photon flux $\Gamma_{v}$ in Eq.~\eqref{eq:weight_cr_sect}. In that case the weight corresponds to the hadronic cross section under the influence of virtual photons. This mode is convenient   if one wants to unfold the hadronic cross section value from the EG distributions (see Sect.~\ref{unfold} for details).

\chapter{Unfolding the cross section value from the EG distributions}
\label{unfold}



As documented in Sect.~\ref{sect:data}, the value of the seven-differential double pion cross section is applied as a weight $f_{cr~sect}$ to each generated event. The EG distributions, where generated events are summed up with weights, can serve for the purpose of unfolding the values of integrated and single-differential cross sections. In order to do this, a statistically sufficient number of events needs to be generated in the $F_{flux}=0$ mode. Let's consider separately the issues of unfolding the values of integral and single-fold differential cross sections.

\begin{enumerate}

\item Unfolding the integral cross section value from the EG distributions.

Let's assume that $N_{evt}^{\Delta W,\Delta Q^2}$ events are generated in a particular $W$ and $Q^2$ bin with the widths $\Delta W$ and $\Delta Q^2$, respectively. The EG yield $Y_{gen}$ is a weighted sum of these events,

\begin{equation}
Y_{gen} (\Delta W,\Delta Q^2)= \sum_{i=1}^{N_{evt}^{\Delta W,\Delta Q^2}} f_{cr~sect}^{i}  = \sigma(\Delta W,\Delta Q^2)N_{evt}^{\Delta W,\Delta Q^2},
\label{eq:unfold_int_1}
\end{equation}
where $\sigma(\Delta W,\Delta Q^2)$ is the desired value of the integral cross section in $W$ and $Q^2$ bin, which should be unfolded.

Therefore, in order to obtain the value of the integral cross section in a particular $W$ and $Q^2$ bin, one needs to scale the weighted sum of generated events with the number of events in that bin,

\begin{equation}
\sigma(\Delta W,\Delta Q^2)=\frac{Y_{gen} (\Delta W,\Delta Q^2)}{N_{evt}^{\Delta W,\Delta Q^2}}.\\
\label{eq:unfold_int_2}
\end{equation}

It should be mentioned that Eq.~\eqref{eq:unfold_int_1} and Eq.~\eqref{eq:unfold_int_2} are only relevant for a large number of generated events, because the summation over all events implies also the summation over all final hadron variables, since the weight represents the seven-dimensional cross section. A single point in $W$ and $Q^2$ corresponds to the range in each final hadron variable. Therefore, in order to obtain the correct value of the integral cross section, these ranges must be well-populated with events. So,  Eq.~\eqref{eq:unfold_int_2} will not lead to the correct value of $\sigma(\Delta W,\Delta Q^2)$ if just few events are generated.

The simplicity of the relations Eq.~\eqref{eq:unfold_int_1} and Eq.~\eqref{eq:unfold_int_2} is a consequence of the fact that proper normalization factors are embedded in the weight $f_{cr~sect}$, as Eq.~\eqref{eq:weight_cr_sect} demonstrates. These factors were chosen based on the demand that the weighted sum of the events must lead to  the integral cross section value.

It is convenient to use one-dimensional ROOT histograms to unfold the cross section from the EG distributions. Here is an example of obtaining the $W$ dependence of integrated cross section for a particular $Q^2$ bin by means of filling the histogram and its subsequent scaling. In order to do so, one should generate $N_{evt}$ in the limits $[W_{min},~W_{max}]$ and $[Q^2_{min},~Q^2_{max}]$. Then the following histogram should be created.

TH1F *h\_w = new TH1F ("h\_w", "h\_w", $n_{bins}^{w}$, $W_{min}$, $W_{max}$);

This histogram should be filled inside the loop over all events that are weighted with $f_{cr~sect}$.

h\_w $\rightarrow$ Fill($W$, $f_{cr~sect}$);

After that the histogram should be scaled with the factor $F$, which corresponds to the number of events inside one bin.

$F = \frac{N_{evt}}{n_{bins}^{w}}$;

This formula is a consequence of the flat event generation and the greater the number of generated events is, the more accurate this relation becomes.

h\_w $\rightarrow$ Scale($1/F$);

Eventually the histogram h\_w contains the $W$ dependence of the integrated cross section (in $\mu b$) in the limits $[W_{min},~W_{max}]$ for the $Q^2$ bin $[Q^2_{min},~Q^2_{max}]$.

\item Unfolding the single-differential cross section value from the EG distributions.

Let's again assume that $N_{evt}^{\Delta W,\Delta Q^2}$ events are generated in a particular $W$ and $Q^2$ bin with the widths $\Delta W$ and $\Delta Q^2$, respectively. If one is interested in the cross section that is single-differential in the final hadron variable $X$ in the bin $\delta X$ with the number of events $N_{evt}^{\Delta W,\Delta Q^2, \delta X}$, then firstly the following weighted sum should be considered.

\begin{equation}
\begin{aligned}
Y_{gen}(\Delta W,\Delta Q^2,\delta X)&= \sum_{i=1}^{N_{evt}^{\Delta W,\Delta Q^2, \delta X}}f_{cr~sect}^{i} = N_{evt}^{\Delta W,\Delta Q^2, \delta X}\frac{d\sigma}{dX} (\Delta W,\Delta Q^2,\delta X)\Delta X = \\
&= \left [ N_{evt}^{\Delta W,\Delta Q^2 }\frac{\delta X}{\Delta X}\right ] \left [\frac{d\sigma}{dX} (\Delta W,\Delta Q^2,\delta X)\Delta X\right ] = \\
&= \frac{d\sigma}{dX} (\Delta W,\Delta Q^2,\delta X)N_{evt}^{\Delta W,\Delta Q^2}\delta X, 
\label{eq:first_case}
\end{aligned}
\end{equation}

where $\frac{d\sigma}{dX}(\Delta W,\Delta Q^2,\delta X)$  is the desired value of the single-differential cross section in $\Delta W$, $\Delta Q^2$ and $\delta X$ bin, which should be unfolded. $\Delta X$ is the full range kinematically available for the variable $X$. The fraction in the first square brackets appears, because one is interested in the amount of events in $\delta X$ bin, which in case of flat generation is connected with the total number of events in $\Delta W$, $\Delta Q^2$ bin by this fraction. The multiplier $\Delta X$ in the second square brackets appears, because according to Eq.~\eqref{eq:weight_cr_sect} the weight contains the normalization factors, which are equal to the full ranges in different final hadron variables and are needed to force the weights to give the proper integral cross section value upon the weighted summation of all events in the $\Delta W$, $\Delta Q^2$ bin. Since in this case the events are summed over all final variables, except of the variable $X$, the full range $\Delta X$ appears in the brackets. 

Therefore, in order to obtain the value of the single-differential cross section in a particular $\Delta W$, $\Delta Q^2$, and $\delta X$ bin the following scaling of the generated yield in this bin should be performed.

\begin{equation}
\frac{d\sigma}{dX} (\Delta W,\Delta Q^2,\delta X) = \frac{Y_{gen}(\Delta W,\Delta Q^2,\delta X)}{N_{evt}^{\Delta W,\Delta Q^2}\delta X} =  \frac{Y_{gen}(\Delta W,\Delta Q^2,\delta X)}{N_{evt}^{\Delta W,\Delta Q^2,\delta X}\Delta X}
\end{equation}

In terms of one-dimensional ROOT histograms the unfolding procedure is described next. Let's assume that $N_{evt}$ events are generated in one $\Delta W$, $\Delta Q^2$ bin.

\begin{itemize}
\item For the invariant mass distributions the following steps should be carried out.

TH1F *h\_m = new TH1F ("h\_m", "h\_m", $n_{bins}^{M}$, $M_{min}$ , $M_{max}$);

h\_m $\rightarrow$ Fill(M, $f_{cr~sect}$);

$F = \frac{N_{evt}(M_{max}-M_{min})}{n_{bins}^{M}}$;

h\_m $\rightarrow$ Scale($1/F$);

After that the histogram h\_m contains the single-differential cross section $\frac{d\sigma}{dM}$ in $\mu{\rm b/GeV}$.

\item For the $\alpha_{h}$ and $\varphi_{h}$ angular distributions the following steps should be carried out.

TH1F *h\_ang = new TH1F ("h\_ang", "h\_ang", $n_{bins}^{ang}$, 0, $2\pi$);

h\_ang $\rightarrow$ Fill(ang, $f_{cr~sect}$);

$F = \frac{2\pi N_{evt}}{n_{bins}^{ang}}$

h\_ang $\rightarrow$ Scale($1/F$);

After that the histogram h\_ang contains the single-differential cross section $\frac{d\sigma}{d\alpha_{h}}$ or $\frac{d\sigma}{d\varphi_{h}}$  in $\mu{\rm b/rad}$.

\item For the $\frac{d\sigma}{d(-cos\theta_{h})}$ single-differential cross section the procedure is a little bit more complicated. Two one-dimensional histograms should be created first.

TH1F *h1 = new TH1F("h1","h1", $n_{bins}^{th}$, 0 , $\pi$); 
 
TH1F *h2 = new TH1F("h2","h2", $n_{bins}^{th}$, 0 , $\pi$);

The histogram h1 should be filled inside the loop over all events.

h1 $\rightarrow$ Fill($\theta_{h}$, $f_{cr~sect}$);

After h1 is filled, the content of the histogram h2 should be set by the following way.

 for (Short\_t i=1; ${\rm i}\leq n_{bins}^{th}$; i++)\{

$\delta_{cos}$ = cos(h1$\rightarrow$GetBinLowEdge(i)) - cos(h1$\rightarrow$GetBinLowEdge(i)+h1$\rightarrow$GetBinWidth(i));

h2$\rightarrow$  SetBinContent (i, h1 $\rightarrow$ GetBinContent(i)/$\delta_{cos}$);

     \};

h2 $\rightarrow$ Scale($1/N_{evt}$);

 After that the histogram h2 contains the single-differential cross section $\frac{d\sigma}{d(-cos\theta_{h})}$  in $\mu{\rm b/rad}$.

\end{itemize}

\end{enumerate}

\chapter{Quality of the data description}
\label{quality}
As described in Sect.~\ref{unfold}, one can simply unfold the values of the integrated and single-differential cross sections from the EG distributions.  This section presents the plots that illustrate how well the EG describes the input data in the regions, where they exist. All TWOPEG distributions, which are shown below, have been generated in the $F_{flux}=0$ mode.  

\begin{figure}[!ht]
\begin{center}
\includegraphics[width=0.75\textwidth]{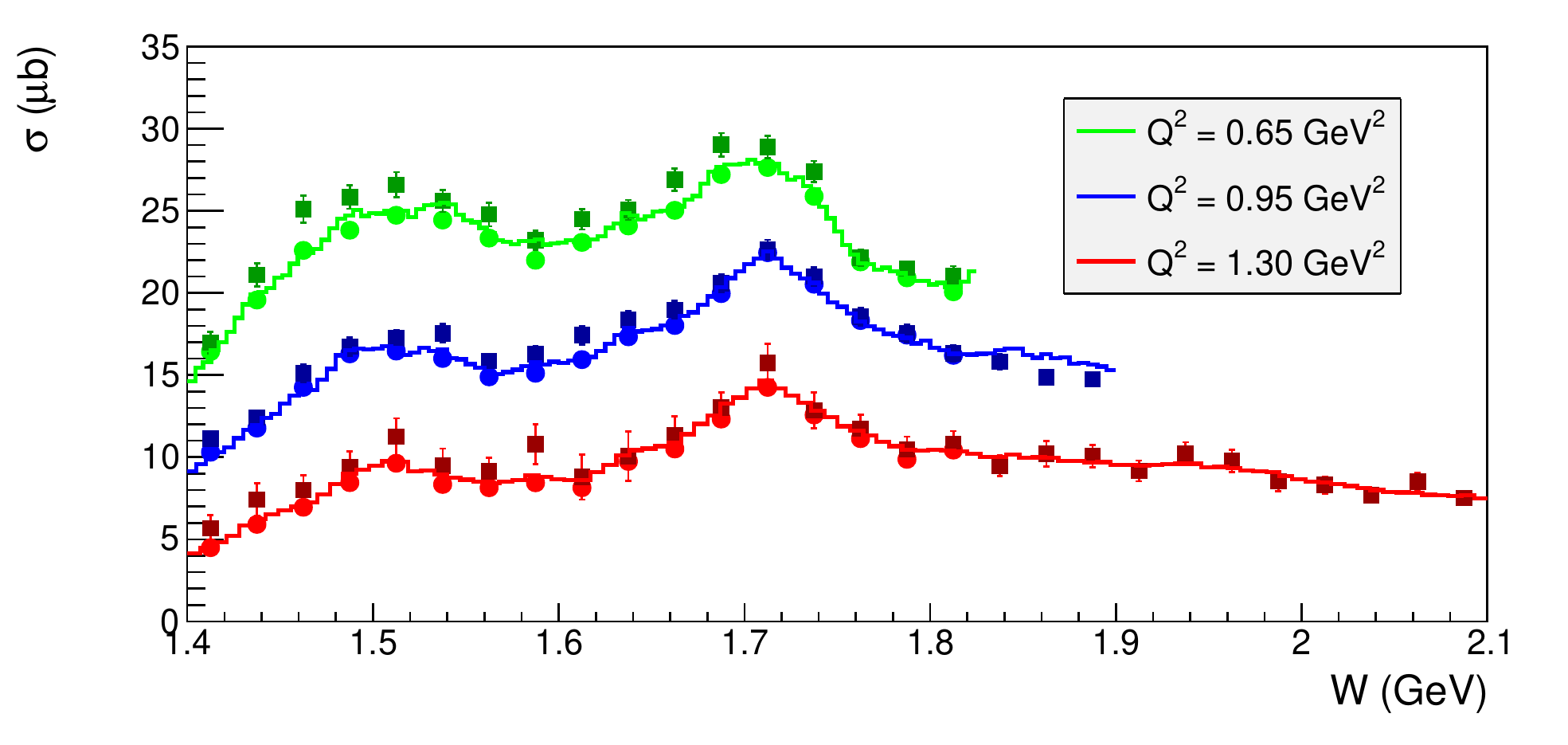}
\end{center}
\vspace{-0.6cm}
\caption{\small Comparison between event distributions of TWOPEG (curves) for different $Q^2$ bins, the integrated cross sections from the JM model~\cite{Mokeev:2015lda} (circles), and data~\cite{Ripani:2002ss} (squares) for the corresponding three $Q^2$ points at 0.65, 0.95, and 1.3~GeV$^2$. To obtain the green and blue curves the beam energy was set to 2.445 GeV, while for the red one the beam energy was set to 4 GeV. This comparison corresponds to the data-set 1a, which is marked in Fig.~\ref{fig:gen_cover} as the region within the red boundaries. }
\label{fig:eg_rip}
\end{figure}

\begin{figure}[!ht]
\begin{center}
\includegraphics[width=0.6\textwidth]{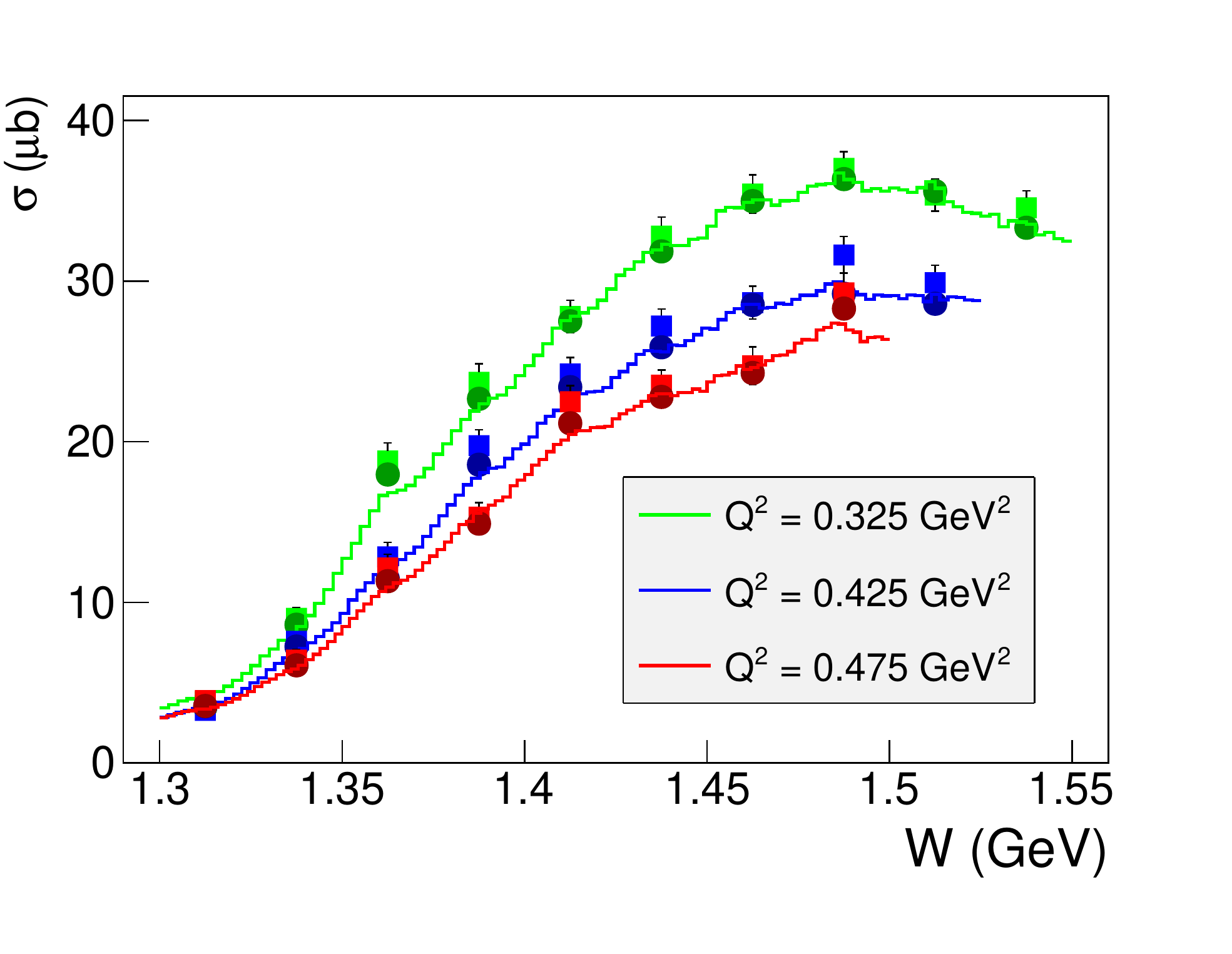}
\end{center}
\vspace{-0.6cm}
\caption{\small Comparison of the TWOPEG event distributions (curves) for different $Q^2$ bins with the integrated cross sections from the JM model~\cite{Mokeev:2008iw},~\cite{Mokeev:2012vsa} (circles) and data~\cite{Fedotov:2008aa} (squares) for the corresponding three $Q^2$ points at 0.325, 0.425, and 0.475~GeV$^2$. The TWOPEG distributions were obtained for $E_{beam} = 1.515$ GeV. This comparison corresponds to the data-set 1b, which is marked in Fig.~\ref{fig:gen_cover} as the region within the lilac boundaries.}
\label{fig:eg_fed}
\end{figure}

\begin{figure}[!ht]
\begin{center}
\frame{
\includegraphics[height=0.45\textwidth]{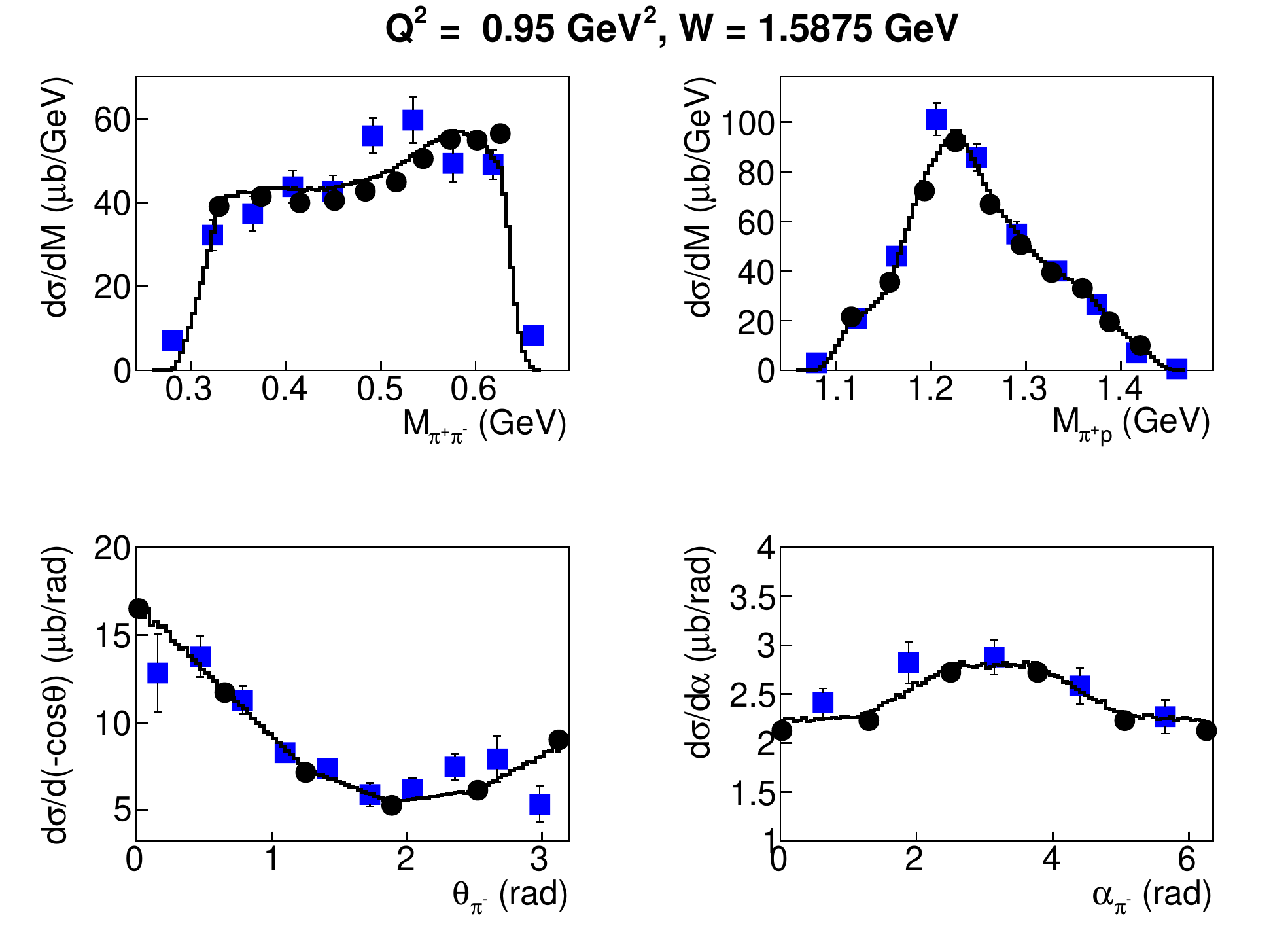}
}
\end{center}
\vspace{-0.6cm}
\caption{\small Comparison of the TWOPEG event distributions (curves) for the $W$ bin [1.575,~1.6]~GeV and $Q^2$ bin [0.8,~1.1]~GeV$^2$ with the single-fold differential cross sections from the JM model~\cite{Mokeev:2015lda} (circles) and data~\cite{Ripani:2002ss} (squares) for the $W = 1.5875$~GeV, $Q^2 = 0.95$~GeV$^2$ point. The TWOPEG distributions were obtained for $E_{beam} = 2.445$ GeV. This comparison corresponds to the data-set 1a, which is marked in Fig.~\ref{fig:gen_cover} as the region within the red boundaries.}
\label{fig:eg_rip_095_15875}
\end{figure}

\begin{figure}[!ht]
\begin{center}
\frame{
\includegraphics[height=0.45\textwidth]{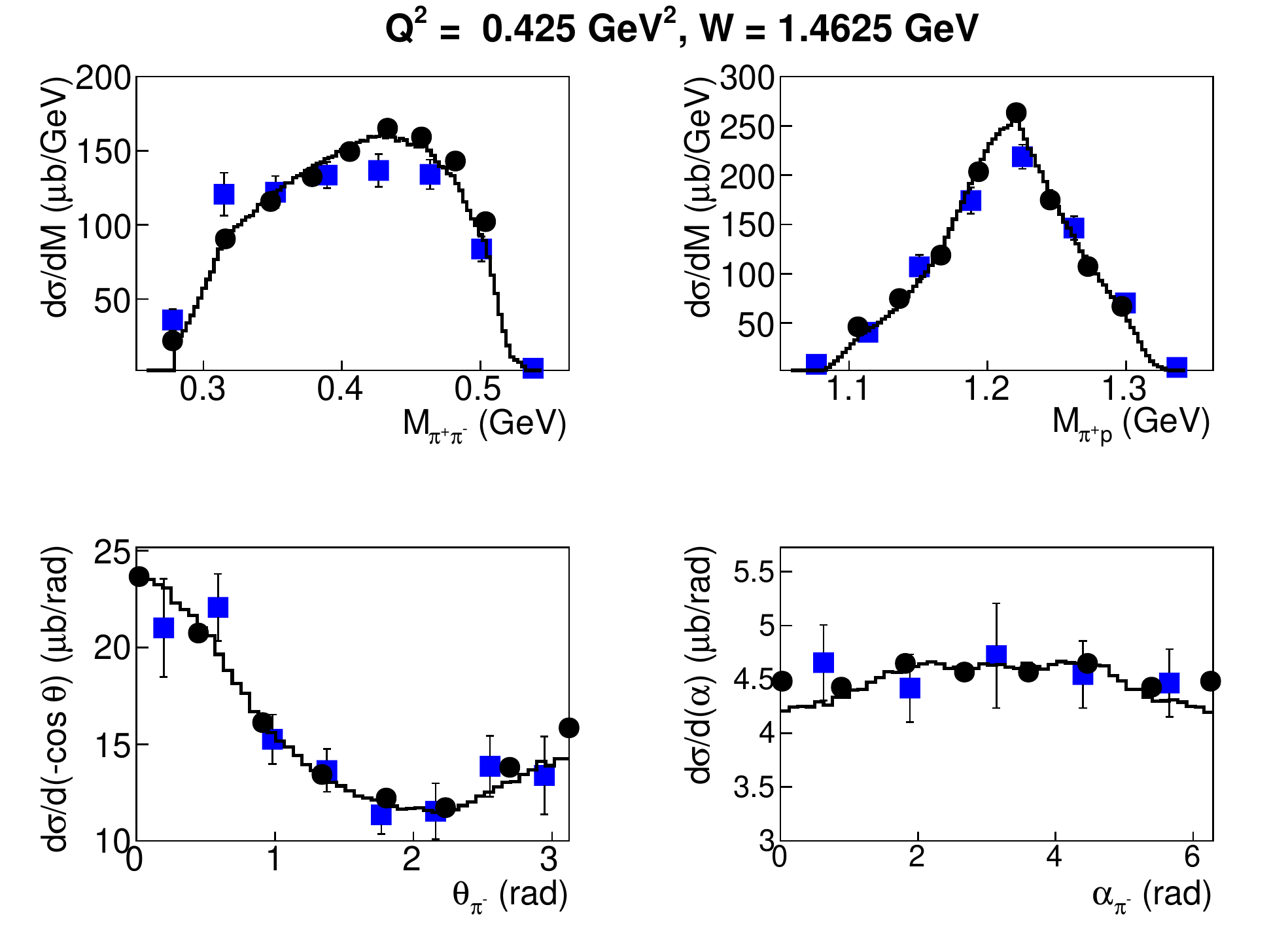}
}
\end{center}
\vspace{-0.6cm}
\caption{\small Comparison of the TWOPEG event distributions (curves) for the $W$ bin [1.45,~1.475]~GeV and $Q^2$ bin [0.4,~0.45]~GeV$^2$ with the single-fold differential cross sections from the JM model~\cite{Mokeev:2008iw},~\cite{Mokeev:2012vsa} (circles) and data~\cite{Fedotov:2008aa} (squares) for the $W = 1.4625$~GeV, $Q^2 = 0.425$~GeV$^2$ point. The TWOPEG distributions were obtained for $E_{beam} = 1.515$ GeV. This comparison corresponds to the data-set 1b, which is marked in Fig.~\ref{fig:gen_cover} as the region within the lilac boundaries.}
\label{fig:eg_fed_0425_14625}
\end{figure}

Figures~\ref{fig:eg_rip} and \ref{fig:eg_fed} show the comparison between the event distributions of TWOPEG (curves) with the integrated cross sections from the JM model~\cite{Mokeev:2015lda},~\cite{Mokeev:2008iw},~\cite{Mokeev:2012vsa}  (circles) and measured data~\cite{Ripani:2002ss},~\cite{Fedotov:2008aa} (squares) for different values of $Q^2$ for the electroproduction data-sets 1a and 1b, respectively. In Figs.~\ref{fig:eg_rip_095_15875} and \ref{fig:eg_fed_0425_14625} the analogous comparison is made for the single-differential cross sections of these data-sets.

\begin{figure}[!ht]
\begin{center}
\frame{
\includegraphics[height=0.45\textwidth]{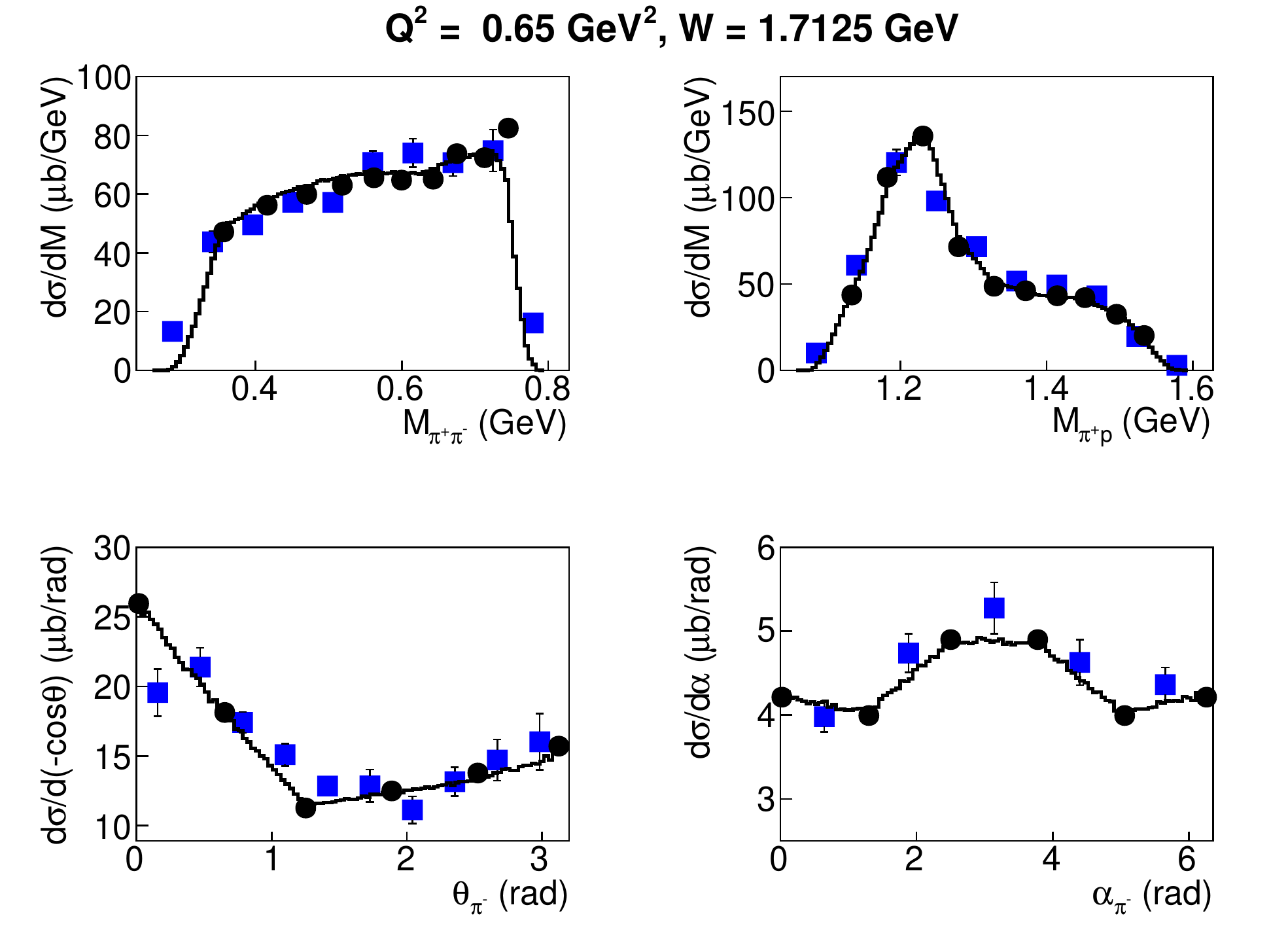}
}
\end{center}
\vspace{-0.6cm}
\caption{\small Comparison of the TWOPEG event distributions (curves) for the $W$ bin [1.7,~1.725]~GeV and $Q^2$ bin [0.5,~0.8]~GeV$^2$ with the single-fold differential cross sections from the JM model~\cite{Mokeev:2015lda} (circles) and data~\cite{Ripani:2002ss} (squares) for the $W = 1.7125$~GeV, $Q^2 = 0.65$~GeV$^2$ point. The TWOPEG distributions were obtained for $E_{beam} = 2.445$ GeV. This comparison corresponds to the data-set 1a, which is marked in Fig.~\ref{fig:gen_cover} as the region within the red boundaries.}
\label{fig:eg_rip_065_17125}
\end{figure}

\begin{figure}[!ht]
\begin{center}
\frame{
\includegraphics[height=0.45\textwidth]{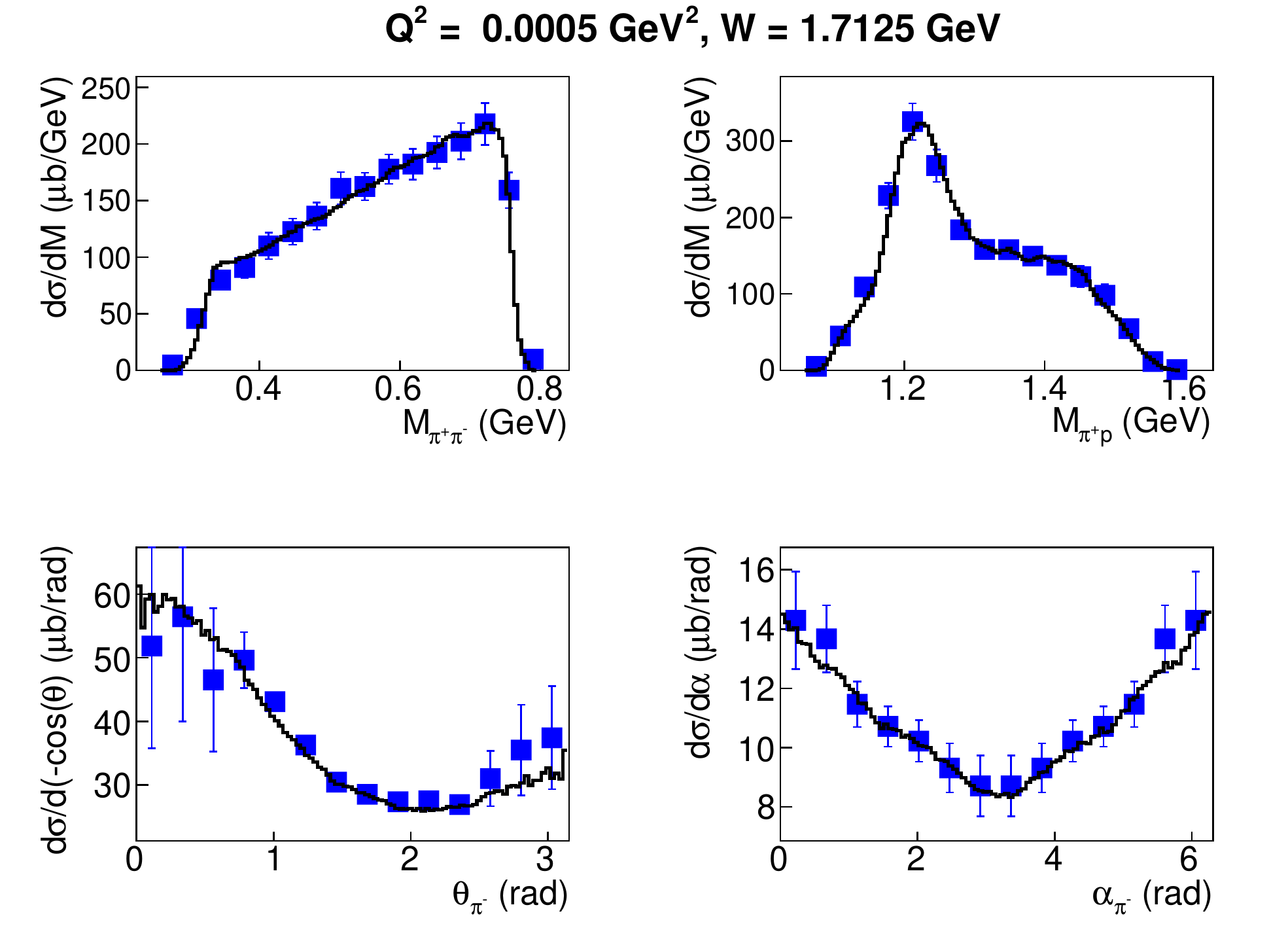}

}
\end{center}
\vspace{-0.6cm}
\caption{\small Comparison of the TWOPEG event distributions (curves) for the $W$ bin [1.7,~1.725]~GeV and $Q^2$ bin [0.0004,~0.0006]~GeV$^2$ with the single-fold differential experimental cross sections~\cite{Golovach:note} (squares) for the $W = 1.7125$~GeV, $Q^2 = 0$~GeV$^2$ point. This comparison corresponds to the data-set 2a, which is marked in Fig.~\ref{fig:gen_cover} by the green line.}
\label{fig:eg_gol_17125}
\end{figure}

The comparison that is shown in Figs.~\ref{fig:eg_rip_065_17125} and \ref{fig:eg_gol_17125}, corresponds to the $Q^2$ boundaries of the region II in Fig.~\ref{fig:gen_cover}. As it is described in Sect.~\ref{sect:cr_sect_extr_intr}, the five-dimensional cross section in this region is a mixture of the two cross section samples, each scaled from the corresponding $Q^2$ edge at $Q^2 = 0$~GeV$^2$ and $Q^2 = 0.65$~GeV$^2$. The mixing is made in the way that at the $Q^2$ edge the resultant cross section coincides with the corresponding model cross section at this edge (see Eq.~\eqref{eq:q2_dep_fit_sig_t}). Figure~\ref{fig:eg_rip_065_17125} shows the comparison between event distributions of TWOPEG and the single-differential model cross sections at the $Q^2 = 0.65$~GeV$^2$ edge, while Fig.~\ref{fig:eg_gol_17125} shows the same for the $Q^2 = 0$~GeV$^2$ edge. 

\begin{figure}[!ht]
\begin{center}
\frame{
\includegraphics[height=0.45\textwidth]{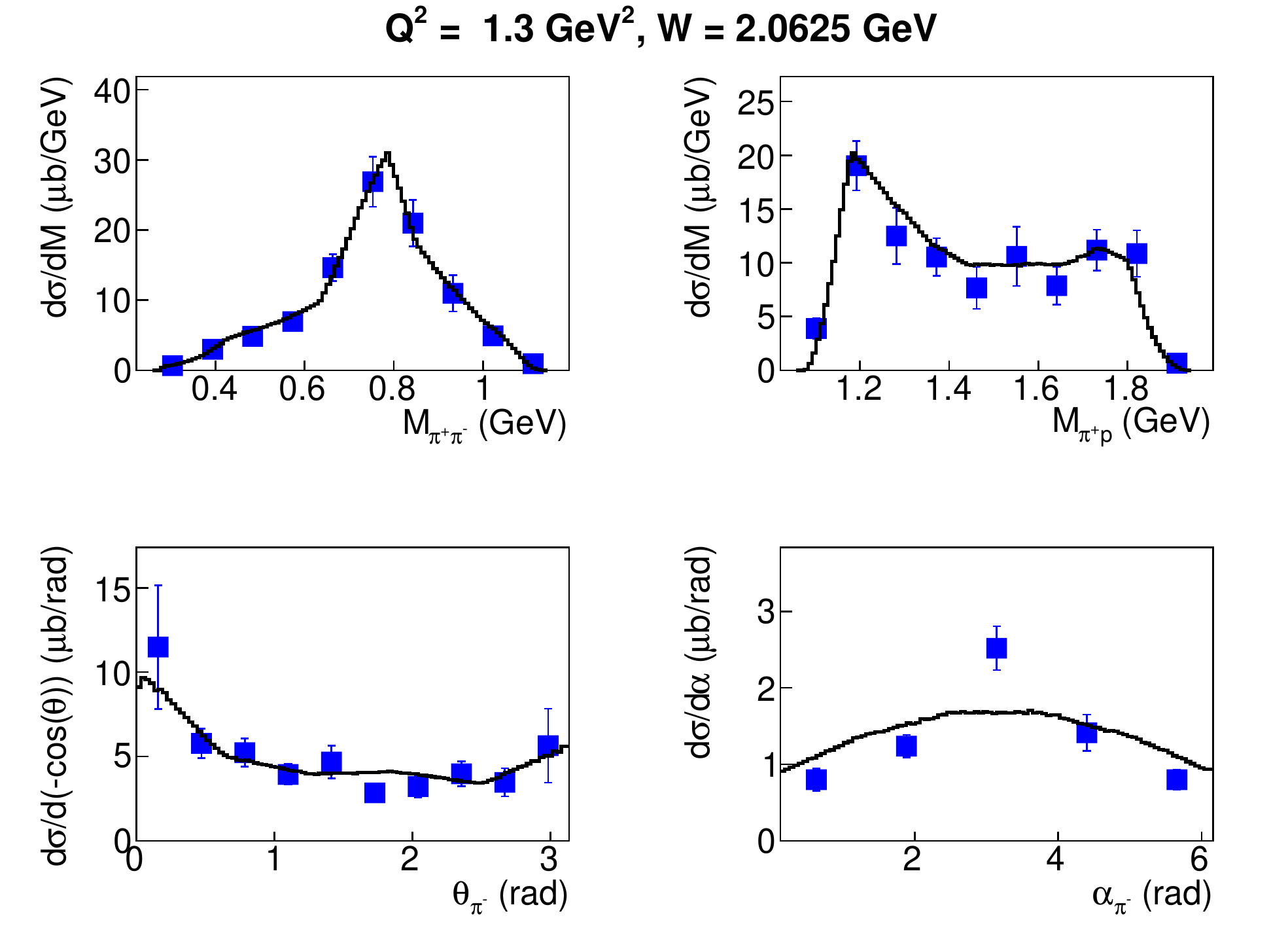}

}
\end{center}
\vspace{-0.6cm}
\caption{\small Comparison of the TWOPEG event distributions (curves) for the $W$ bin [2.05,~2.075]~GeV and $Q^2$ bin [1.25,~1.35]~GeV$^2$ with the single-fold differential experimental cross sections~\cite{Ripani:2002ss} (squares) for the $W = 2.0625$~GeV, $Q^2 = 1.3$~GeV$^2$ point. The TWOPEG distributions were obtained for $E_{beam} = 4$ GeV. This comparison corresponds to the red line at $Q^2 = 1.3$~GeV$^2$, that is adjacent to the data-set 1a in Fig.~\ref{fig:gen_cover}.}
\label{fig:eg_rip_130_20625}
\end{figure}

\begin{figure}[!ht]
\begin{center}
\frame{
\includegraphics[height=0.45\textwidth]{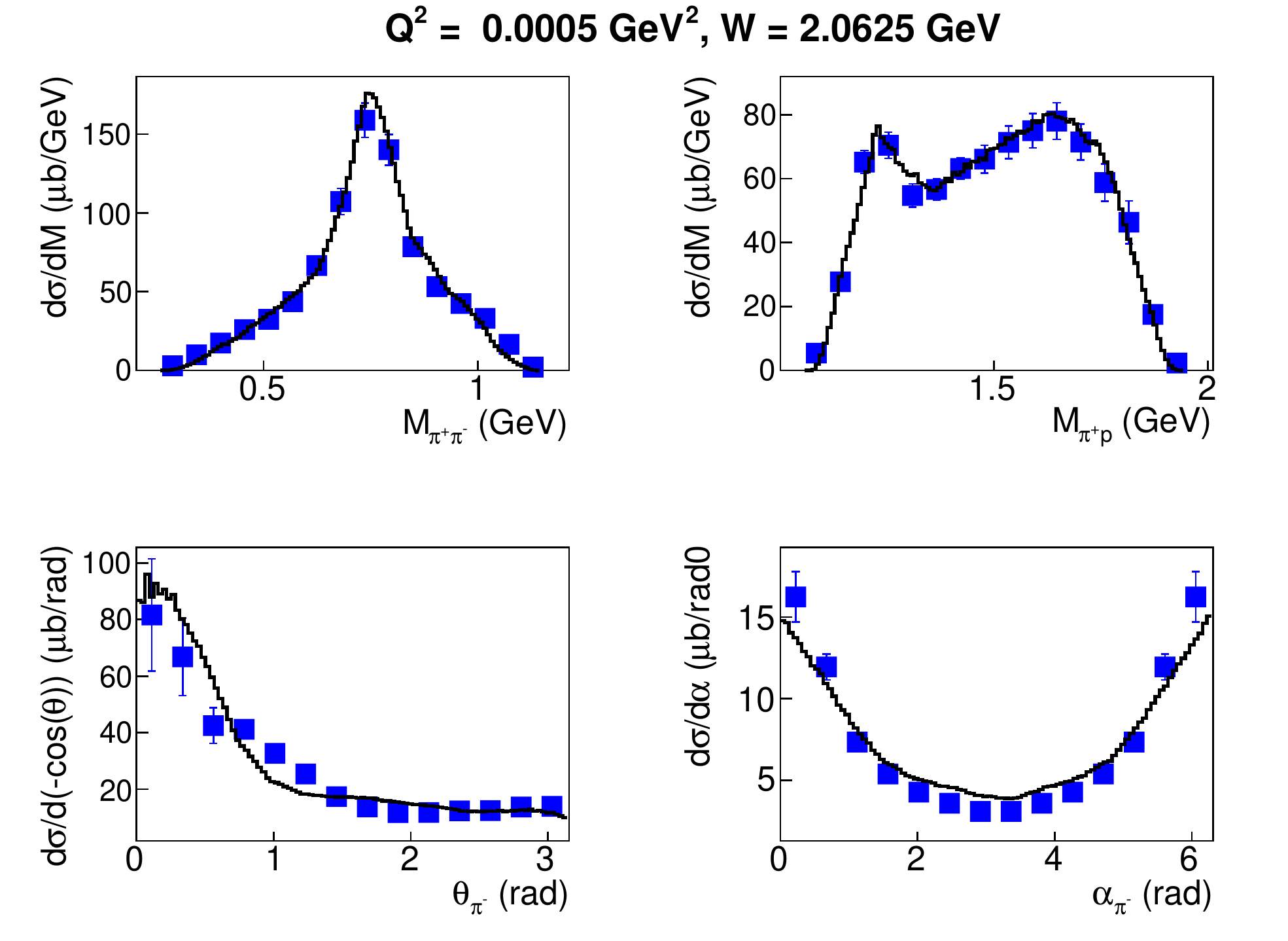}

}
\end{center}
\vspace{-0.6cm}
\caption{\small Comparison of the TWOPEG event distributions (curves) for the $W$ bin [2.05,~2.075]~GeV and $Q^2$ bin [0.0004,~0.0005]~GeV$^2$ with the single-fold differential experimental cross sections~\cite{Golovach:note} (squares) for the $W = 2.0625$~GeV, $Q^2 = 0$~GeV$^2$ point. This comparison corresponds to the data-set 2a, which is marked in Fig.~\ref{fig:gen_cover} by the green line.}
\label{fig:eg_gol_20625}
\end{figure}

The comparison that is shown in Figs.~\ref{fig:eg_rip_130_20625} and \ref{fig:eg_gol_20625}, corresponds to the $Q^2$ boundaries of the region IV in Fig.~\ref{fig:gen_cover}. The same idea of cross section mixing is applied here. Figure~\ref{fig:eg_rip_130_20625} shows the comparison between TWOPEG event distributions and the single-differential experimental cross sections at the $Q^2 = 1.3$~GeV$^2$ edge, while Fig.~\ref{fig:eg_gol_20625} shows the same for the $Q^2 = 0$~GeV$^2$ edge. 

\begin{figure}[!ht]
\begin{center}
\includegraphics[width=0.65\textwidth]{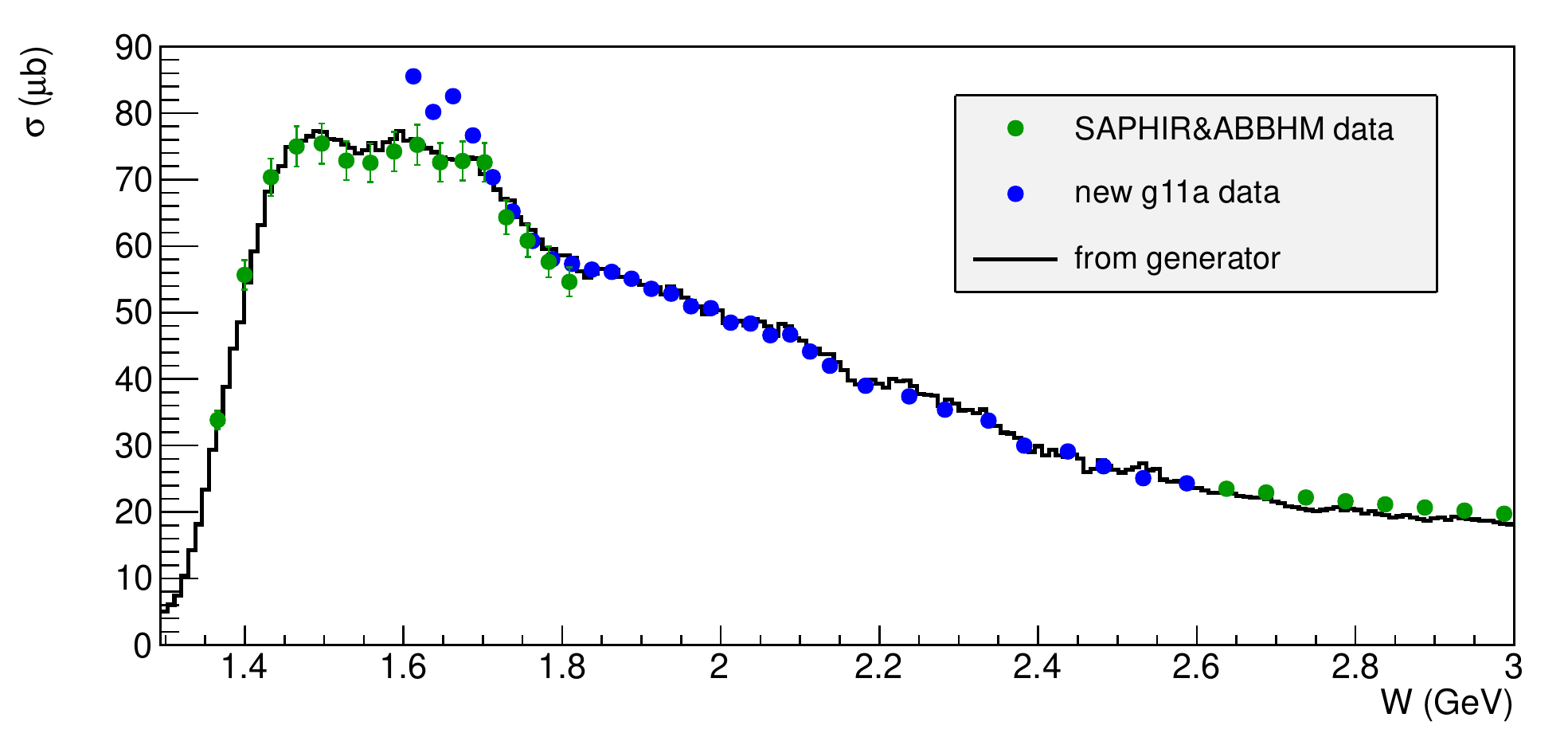}
\includegraphics[width=0.65\textwidth]{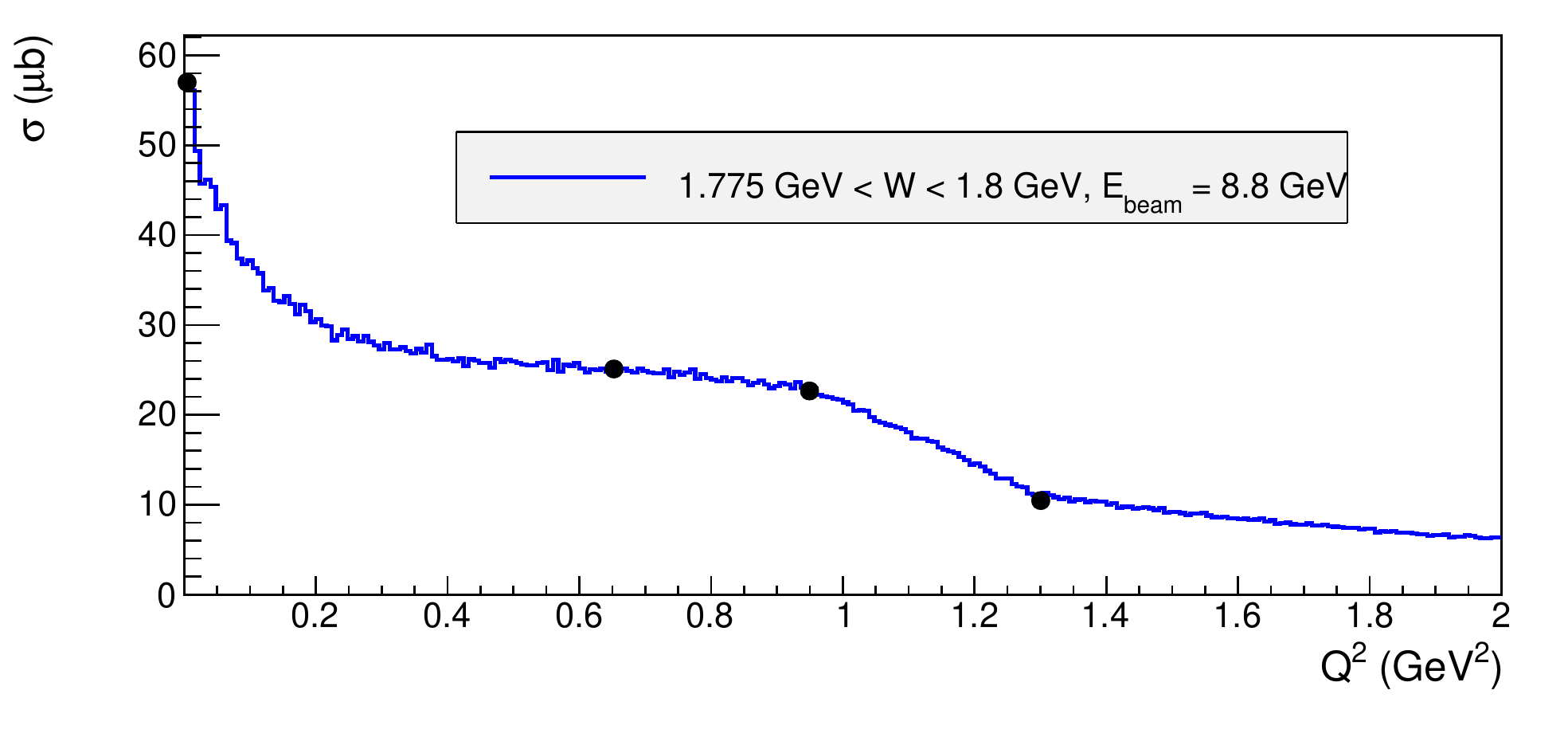}
\end{center}
\vspace{-0.6cm}
\caption{\small Upper plot shows the $W$ dependence of the integrated cross section for quasi-real photons with $Q^2$ [0.0004,~0.0005]~GeV$^2$ in comparison with data~\cite{Golovach:note,Wu:2005wf,ABBHHM:1968aa}. Lower plot shows a typical example of $Q^2$ dependence of the total cross section for one $W$ bin in comparison with the JM model~\cite{Mokeev:2015lda}  at $W = 1.7875$~GeV for a beam energy of 8.8~GeV.}
\label{fig:eg_ph_point}
\end{figure}

Figure~\ref{fig:eg_ph_point} (upper plot) demonstrates that TWOPEG nicely reproduces the behavior of the total integrated cross section close to $Q^2 = 0$~GeV$^2$. Furthermore, the lower plot in Fig.~\ref{fig:eg_ph_point} shows a typical example of the $Q^2$ dependence of the total cross section. 

In any desired point of the regions 1a, 1b, and 2a in Fig.~\ref{fig:gen_cover} TWOPEG produces the established values of the model cross section. In the regions I, II, and IV the EG gives some cross section estimation based on the cross section in the neighboring areas and can therefore be used as a prediction of a naive model. In other regions the cross section estimates are less reliable, but good enough for the modeling of experiments (efficiency evaluation, background estimation, etc.).  


\clearpage
\newpage

\chapter{Simulation of radiative effects}
\label{rad_eff}

In order to simulate radiative effects (RE) the Mo and Tsai approach~\cite{Mo:1968cg} was chosen. This approach allows to calculate radiative integral cross sections from the given nonradiative ones in each $(W,~Q^{2})$ point. In~\cite{Mo:1968cg} this is applied to the inclusive case, while here double pion integral cross sections are used instead. 

It is essential that approach~\cite{Mo:1968cg} only accounts for the change of cross section values, therefore an additional procedure (discussed below) is used to generate the energy of the radiative photon and to account for the shift in $W$ and $Q^2$ caused by RE.

It also needs to be mentioned that approach~\cite{Mo:1968cg} assumes that radiative photons are emitted colliniarly to the in- and outgoing electron directions (so-called "peaking approximation"). The minimal energy of the emitted radiative photon is a free parameter. It is denoted as $\Delta$ and chosen to be equal to 10~MeV. In~\cite{Mo:1968cg} it is claimed that the result was found to be insensitive to the choice of $\Delta$ value, however this value must be smaller than the resolution of the experiment.  


As it is derived in~\cite{Mo:1968cg} (Eq. (IV.1) on page 213) the radiative cross section in each ($W$,~$Q^2$) point is given by~\footnote[1]{Note that~\cite{Mo:1968cg} assumes the cross section to be differential in ($\Omega$,~$E_{e'}$), while TWOPEG assumes it to be differential in ($W$,~$Q^2$). Therefore, the nonradiative hadronic cross section is taken for the needed ($W$,~$Q^2$) point and then multiplied by the virtual photon flux $\Gamma_{v} = \frac{\alpha}{4\pi^2}\frac{E_{e'}}{E_{beam}m_{p}}\frac{(W^{2}-m_{p}^{2})}{(1-\varepsilon_{T})Q^{2}}$, which is connected with the one defined by~Eq.\eqref{flux} via the Jacobian for the ($W$,~$Q^2$)$\rightarrow$($\Omega$,~$E_{e'}$) coordinate transformation.}

\begin{equation}
\frac{d\sigma_{rad}}{d\Omega  dE_{e'}}(W,Q^{2}) = S_{1} + S_{2} + S_{3}.
\label{eq:rad_tot}
\end{equation}

Three terms in Eq.~\eqref{eq:rad_tot} correspond to various regions of integration in Fig.~3 on page 216 of~\cite{Mo:1968cg}. Contribution from region four in this figure is neglected in order to save computation time.

The term $S_{1}$ corresponds to the so-called "soft radiation", in which the energy of the radiated photon is less than $\Delta$. The two remaining terms correspond to the so-called "hard radiation", in which the energy of the radiated photon is greater than $\Delta$. $S_{2}$ accounts for the changing initial electron energy and $S_{3}$ for the changing scattered electron energy.

Let's look at each term of Eq.~\eqref{eq:rad_tot} in more detail.
\begin{enumerate}

\item $S_{1}$ can be factorized in the following way~\footnote[2]{The Eq.~\eqref{eq:rad_s1} is coded in the function {\em s1\_radsoft} and embedded in {\em radcorr.cxx}.} (see Eq.~(IV.1) in~\cite{Mo:1968cg}).
\begin{equation}
S_{1}(W,Q^2) = \frac{d\sigma_{norad}}{d\Omega  dE_{e'}}(W,Q^2)\cdot R_{soft} = \frac{d\sigma_{norad}}{d\Omega  dE_{e'}}(W,Q^2)\cdot e^{\delta_{t}+\delta_{r}},
\label{eq:rad_s1}
\end{equation} 
where $\delta_{t}$ corresponds to the straggling in the target medium, while $\delta_{r}$ corresponds to the radiative correction to continuum spectrum, both of them are defined on page~214 in~\cite{Mo:1968cg}.

Hence the value of $S_1$ at a given ($W$,$Q^2$) point is determined by the nonradiative cross section taken exactly at this point and multiplied by the factor $R_{soft}$.

\item $S_{2}$ is given by~\footnote[3]{The Eq.~\eqref{eq:rad_s2} is coded in the function {\em s2\_radhardini} and embedded in {\em radcorr.cxx}.}
\begin{equation}
\begin{aligned}
S_{2}(W,Q^2) = \int\limits_{\Delta}^{\omega_{max}^{ini}}\rho_{ini}\frac{d\sigma_{norad}}{d\Omega  dE_{e'}}(\widetilde{W},\widetilde{Q^{2}})d\omega, \\
\omega_{max}^{ini} = E_{beam} - \frac{2m_{\pi}^2 + 2m_{p}m_{\pi} + m_{p}E_{e'}}{m_{p}-E_{e'}(1-cos\theta_{e'})},
\label{eq:rad_s2}
\end{aligned}
\end{equation}

where $\omega_{max}^{ini}$ is the maximal energy of the photon that can be emitted by the initial electron assuming two pion production, $\omega$ is the energy of the radiated photon, $\rho_{ini}$ is the part of the integrand defined by Eq.~(IV.1) in~\cite{Mo:1968cg}. $\widetilde{W}$ and $\widetilde{Q^{2}}$ are the shifted values of $W$ and $Q^{2}$ for the given $\omega$ value. $m_{p}$ and $m_{\pi}$ are the proton and charged pion masses, respectively. $E_{beam}$ is the beam energy defined as an input parameter, $E_{e'}$ and $\theta_{e'}$ are the scattered electron energy and polar angle, respectively, given by Eq.~\eqref{eq:el_in_lab}.

It is essential that $S_2$ is calculated numerically as an integral over the radiated photon energy $\omega$, which varies from $\Delta$ to $\omega_{max}^{ini}$ in increments of $\frac{\omega_{max}^{ini}-\Delta}{800}$. Each $\omega$-point of this grid corresponds to the distinct value of the initial electron energy $E_{beam} - \omega$, which in turn corresponds to the shift of the ($\widetilde{W}, \widetilde{Q^{2}}$) values from their initial values. Thus the value of $S_2$ at the point ($W$, $Q^2$) is calculated as an integral of the integrand that is taken for different shifted ($\widetilde{W}, \widetilde{Q^{2}}$) points on this grid.

\item $S_{3}$ is given by~\footnote[4]{The Eq.~\eqref{eq:rad_s3} is coded in the function {\em s3\_radhardfin} and embedded in {\em radcorr.cxx}.}
\begin{equation}
\begin{aligned}
S_{3}(W,Q^2) = \int\limits_{\Delta}^{\omega_{max}^{fin}}\rho_{fin}\frac{d\sigma_{norad}}{d\Omega  d\widetilde{E_{e'}}}(\widetilde{W},\widetilde{Q^{2}})d\omega, \\
\omega_{max}^{fin} = \frac{m_{p}E_{beam} - 2m_{p}m_{\pi} - 2m_{\pi}^2}{m_{p} + E_{beam}(1 - cos\theta_{e'})} - E_{e'},
\label{eq:rad_s3}
\end{aligned}
\end{equation}

where $\omega_{max}^{fin}$ is the maximal energy of the photon that can be emitted by the scattered electron assuming two pion production and $\rho_{fin}$  the part of the integrand defined by Eq.~(IV.1) in~\cite{Mo:1968cg}. 

As in the previous case $S_3$ is calculated numerically as an integral over the radiated photon energy $\omega$, which varies from $\Delta$ to $\omega_{max}^{fin}$ in increments of $\frac{\omega_{max}^{fin}-\Delta}{800}$. Each $\omega$-point of this grid corresponds to the distinct value of the scattered electron energy $\widetilde{E_{e'}} = E_{e'} + \omega$, which in turn corresponds to the ($\widetilde{W}, \widetilde{Q^{2}}$) values shifted from the initial ones. Thus the value of $S_3$ at the point ($W$, $Q^2$) is calculated as an integral of the integrand that is taken for different shifted ($\widetilde{W}, \widetilde{Q^{2}}$) points on this grid. 

\end{enumerate}

The nonradiative integral cross sections, which are needed to compute $S_1$, $S_2$, and $S_3$ are an issue of special attention. The subroutine that simulates RE should have access to the nonradiative integral cross sections for each desired point in $W$ and $Q^2$. Although TWOPEG calculates the cross section value as a weight for each event, it is not justified to use these implemented into the EG cross sections for the purpose of RE simulation. The reason for that is the following: these cross sections are five-differential, therefore, in order to be used for the RE simulation they must firstly be obtained on the five-dimensional grid in final hadron variables and then be integrated over that grid. But for each generated event it would happen more than thousand times, that inevitably leads to an incredible increase in the event generation time.

The following alternative can be used instead: the EG should have direct access to the integral cross sections.
For this purpose, the integrated structure functions $\sigma_{T}$ and $\sigma_{L}$ were obtained from TWOPEG itself (see Sect.~\ref{unfold})  
 on the ($W$,~$Q^2$) grid with bin widths 25~MeV in $W$ and 0.05~GeV$^2$ in $Q^2$ in the limits $[1.2625,\;3.0125]$~GeV in $W$ and $[0.0005,\;1.3]$~GeV$^2$ in $Q^2$, respectively. These structure functions were tabulated in the text files that are located in the folder {\em "int\_sec\_new"}. A two-dimensional linear interpolation is used in order to obtain these structure functions in any given ($W$, $Q^2$) point within the limits described above. Then $\sigma_{T}$ and $\sigma_{L}$ are combined into the full cross section for the given beam energy according to Eq.~\eqref{eq:str_fun_decomp}.
For $W > 3.0125$~GeV the cross section is assumed to be the same as at $W = 3.0125$~GeV for each $Q^2$ point, and for $Q^2 > 1.3$~GeV$^2$ the cross section value at $Q^2 = 1.3$~GeV$^2$ for each $W$ point is scaled with the dependency given by~Eq.~\eqref{eq:q2_dep}.  The interpolation time appears to be negligible in comparison with the time needed to produce the cross section on the five-dimensional grid with the subsequent integration. Therefore, using of this approach simulating of RE increases the event generation time only slightly. 

After the values of $S_1$, $S_2$, and $S_3$ are obtained the following factor can be calculated for each event 
\begin{equation}
f_{rc} (W,Q^2) = \left [S_1+S_2+S_3\right ]/\left [\frac{d\sigma}{d\Omega dE_{e'}}(W,Q^2)\right ].
\end{equation}

This factor shows how the radiative cross section is different from the nonradiative one for each given $W$ and $Q^2$ and is applied as an additional weight factor for each event together with the conventional one $f_{cr~sect}$, which carries information about nonradiative cross section and is discussed in Sect.~\ref{sect:data}.

However, the cross section change due to RE is not the only issue we are interested in. One also wants to simulate the radiative tail in distributions like missing masses, which appears due to the mismatch between the hadron and lepton momenta. This mismatch is the consequence of the fact that ($W$,~$Q^2$) values obtained from the initial and scattered electrons, which suffer from RE, are not those for which final hadrons are produced. To simulate this effect one needs to account for the shift in the ($W$,~$Q^2$) values due to RE, which in turn implies the generation of the radiated photon energy.

To generate the radiated photon energy the random number $R$ is generated in the limits [0,~1]. Then three distinct cases are considered.

\begin{itemize}

\item $0 < R < \frac{S_1}{S_1+S_2+S_3}$

This case corresponds to the so-called "soft" scenario, in which the energy of the radiated photon $E_{rad}$ is less than $\Delta$ and considered to be zero, and ($W$,~$Q^2$) values are considered to be unchanged,

\begin{equation}
\begin{aligned}
& E_{rad} = 0,\\
& \widetilde{Q^2} = Q^2,\\
& \widetilde{W} = W.
\label{eq:first_case}
\end{aligned}
\end{equation}

\item $\frac{S_1}{S_1+S_2+S_3} < R < \frac{S_1+S_2}{S_1+S_2+S_3}$

This case corresponds to the emission of a "hard" photon by the initial electron. The photon energy $E_{rad}$ is generated in the limits [$\Delta$,~$\omega_{max}^{ini}$] according to the probability density~\footnote[5]{Using ROOT it is convenient to do it in the following way.\\
TH1F *h\_radhardini = new TH1F("h\_radhardini","h\_radhardini",800,$\Delta$,$\omega_{max}^{ini}$);\\
for (Int\_t i=0;i$<$800;i++) h\_radhardini$\rightarrow$SetBinContent(i+1,(ARR[i]+ARR[i+1])/2.);\\
R2[0] = hardini\_rndm.Uniform(0.,1.);\\
h\_radhardini$\rightarrow$GetQuantiles(1,eran,R2);\\where ARR[i] is the value of the integrand in Eq.~\eqref{eq:rad_s2} taken on the $\omega$-grid and eran[0] is the photon energy denoted as $E_{rad}$ in the text. } given by the integrand in Eq.~\eqref{eq:rad_s2}. In this case

\begin{equation}
\begin{aligned}
& \widetilde{E}_{beam} = E_{beam} - E_{rad},\\
& \widetilde{Q^2} = 2\widetilde{E}_{beam}E_{e'}(1 - cos\theta_{e'}),\\
& \widetilde{W} = \sqrt{m_{p}^{2} - \widetilde{Q^2} + 2(\widetilde{E}_{beam} - E_{e'})m_{p}},
\label{eq:2nd_case}
\end{aligned}
\end{equation}

where $E_{e'}$ and $\theta_{e'}$ are assumed to be unchanged.

\item $\frac{S_1+S_2}{S_1+S_2+S_3} < R < 1$

This case corresponds to the emission of a "hard" photon by the final electron. The photon energy $E_{rad}$ is generated in the limits [$\Delta$,~$\omega_{max}^{fin}$] according to the probability density given by the integrand in Eq.~\eqref{eq:rad_s3}. In this case

\begin{equation}
\begin{aligned}
& \widetilde{E}_{e'} = E_{e'} + E_{rad},\\
& \widetilde{Q^2} = 2E_{beam}\widetilde{E}_{e'}(1 - cos\theta_{e'}),\\
& \widetilde{W} = \sqrt{m_{p}^{2} - \widetilde{Q^2} + 2(E_{beam} - \widetilde{E}_{e'})m_{p}},
\label{eq:3rd_case}
\end{aligned}
\end{equation}

where $E_{beam}$ and $\theta_{e'}$ are assumed to be unchanged.

\end{itemize}

Now these shifted ($\widetilde{W}$,~$\widetilde{Q^2}$) values are sent as an input to the subroutine that calculates the four-momenta of the final hadrons in the lab frame (see Sect.~\ref{sect:cms_to_lab_trans}). As a result the calculated hadron momenta account for the shift of the ($W$,~$Q^2$) values due to RE. However, the final electron is assumed to be unchanged and still to be defined by Eq.~\eqref{eq:el_in_lab} for the originally generated ($W$,~$Q^2$) values, as well as the initial electron that is assumed to have the unchanged beam energy $E_{beam}$. This corresponds to the real experiment, where one loses information about the change of the lepton momenta due to RE, and the radiative tail in distributions like missing masses appears.

\begin{figure}[!ht]
\begin{center}
\includegraphics[height=0.4\textwidth]{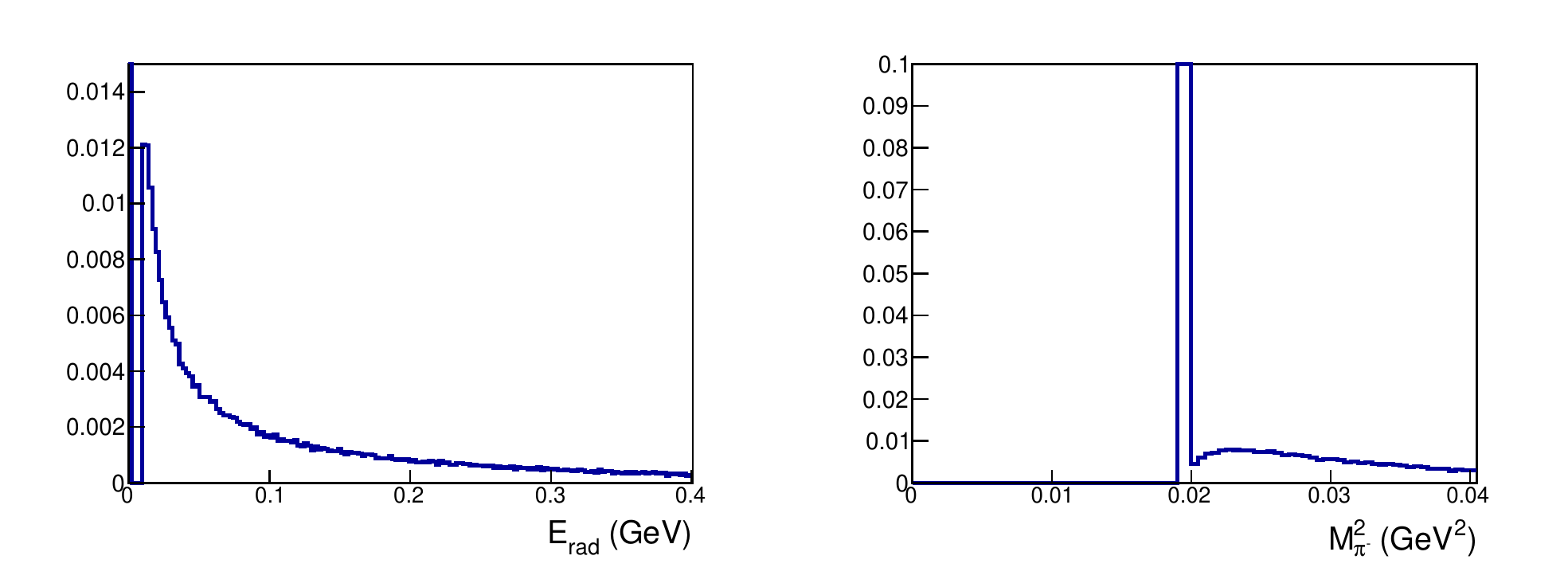}
\end{center}
\vspace{-0.6cm}
\caption{\small Left plot: the distribution of the radiated photon energy $E_{rad}$. Right plot: the distribution of missing mass squared of $\pi^{-}$, which is calculated according to Eq.~\eqref{eq:miss_mass_pim}. Both histograms are normalized in a way that the maxima of the main peaks at zero (left) and at pion mass squared (right) are equal to one. The example is given for the case of $E_{beam} = 2$~GeV, $1.6 < W < 1.8$~GeV, and $0.4 < Q^{2} < 0.5$~GeV$^{2}$.}
\label{fig:miss_mass_pim_eradgam}
\end{figure}

Figure~\ref{fig:miss_mass_pim_eradgam} shows the distributions of the radiated photon energy (left plot) and missing mass squared of the $\pi^{-}$ (right plot). The latter is calculated by
\begin{equation} \label{eq:miss_mass_pim}
M_{\pi^{-}}^{2} = (P_{p} + P_{e} - P_{e'} - \widetilde{P}_{p'} - \widetilde{P}_{\pi^{+}})^{2},
\end{equation}

where $P_{i}$ is the four-momentum of the particle $i$. $\widetilde{P}_{p'}$ and $\widetilde{P}_{\pi^{+}}$ correspond to the hadron momenta calculated for the shifted ($\widetilde{W}$,~$\widetilde{Q^2}$) values, while $P_{e}$ and $P_{e'}$ correspond to the initial unshifted ($W$,~$Q^2$) values. This mimics the conditions of real experiment and leads to the radiative tail, which is clearly seen in the right plot of Fig.~\ref{fig:miss_mass_pim_eradgam}.

\chapter{Building and running TWOPEG}
\label{sect:inp_param}

TWOPEG has two compiling options. 
\begin{itemize}
 
\item "make nobos" compiles without BOS libraries, no output in BOS format is possible in this case.
 
\item "make bos" compiles with BOS libraries. BOS output can be created according to the flag in the input file. Special libraries needed for this option are not available among the standard CLAS libraries on ifarm machines anymore.  So, they were compiled manually and located  at this path "$\sim$gleb/lib/LinuxRHFC8", which is specified in the MakeFile. Since these libraries are not supported, sooner or later they will become irrelevant.
\end{itemize}

\begin{figure}[!ht]
\begin{center}
\includegraphics[width=0.25\textwidth]{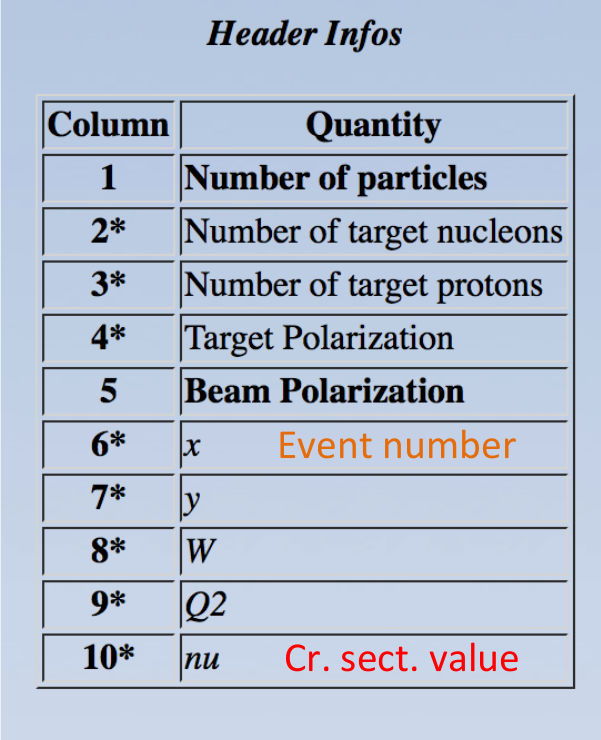}
\end{center}
\vspace{-0.6cm}
\caption{\small The event header in LUND format output. Quantities marked with stars are not used by GEMC, but are still kept in the output stream (user defined meanings could be assigned to them).}
\label{fig:lund_out}
\end{figure}

The compilation and running was tested on ifarm machines.

After the compilation is complete, to run TWOPEG one should type "twopeg\_bos~$<$~inp1" or "twopeg\_nobos $<$ inp1" depending on the compiling options.

The input file "inp1" is located in the EG root directory. It contains input parameters with some comments. In more details these parameters are explained in Tabl.~\ref{tab:input_par}.

\begin{table}[h]\scriptsize
\begin{center}
  \renewcommand{\arraystretch}{1.2}
\begin{tabular}{|c|l|}

\hline
Input parameter & Description\\
\hline 
$N_{evt}$ & Number of events to be generated \\
\hline
$E_{beam}$ & Beam energy (GeV) \\
\hline
$W_{min}$ & W minimum (GeV) \\
\hline
$W_{max}$& W maximum (GeV) \\
\hline
$Q^{2}_{min}$ & $Q^{2}$ minimum (GeV$^2$) \\
\hline
$Q^{2}_{max}$     & $Q^2$ maximum (GeV$^2$) \\
\hline 
$\theta_{min}$ & Minimal $\theta$ of the scattered electron (deg)\\
\hline 
 $\theta_{max}$  & Maximal $\theta$ of the scattered electron (deg)\\
\hline 
$E_{min}$ & Minimal energy of the scattered electron (GeV)\\
\hline 
$R_{targ}$  & Target radius in cm \\
\hline
$L_{targ}$ & Target length in cm \\
\hline
$Z^{off}_{targ}$ & Target offset in $z$ in cm\\
\hline
$\rho_{targ}$ & Target density (g/cm$^3$)\\
\hline
$l^{rad}_{targ}$  & Target radiation length (cm)\\
\hline
$Z_{targ}$ & Target $Z$\\
\hline
$A_{targ}$ & Target $A$\\
\hline
$Th_{wi}$,  $Th_{wf}$   & Thickness of the target windows initial, final (um)\\
\hline
$\rho_{wi}$, $\rho_{wf}$   & Density of target windows initial,final (g/cm$^3$)\\
\hline
$l^{rad}_{wi}$, $l^{rad}_{wf}$  & Radiation length of target windows initial,final (cm)\\
\hline
 &  Output BOS file: 0 - no,\\
$F_{bos}$   &  1 - MCTK,MCVX banks,\\
 &  2 - PART bank\\
\hline
out.bos     & BOS output file name\\
\hline
$F_{lund}$  & Otput LUND file 0 - no, 1 - yes\\
\hline
out.lund    & LUND output file name\\
\hline
 & Radiative mode: 0 - no rad effects,\\
$F_{rad}$   & 1 - rad eff with no straggling, \\ 
 & 2 - rad eff with straggling\\
\hline
$F_{fermi}$  & Fermi smearing: 0 - no, 1 - yes\\
\hline
 &Multiplication by virtual photon flux:\\
$F_{flux}$   &0 - no (under influence of virtual photons),\\
&1 - yes (under influence of electrons)\\
\hline
\end{tabular}
\caption{\small List of the input parameters and their description. \label{tab:input_par}}
\end{center}
\end{table}

\begin{figure}[!ht]
\begin{center}
\includegraphics[width=1.\textwidth]{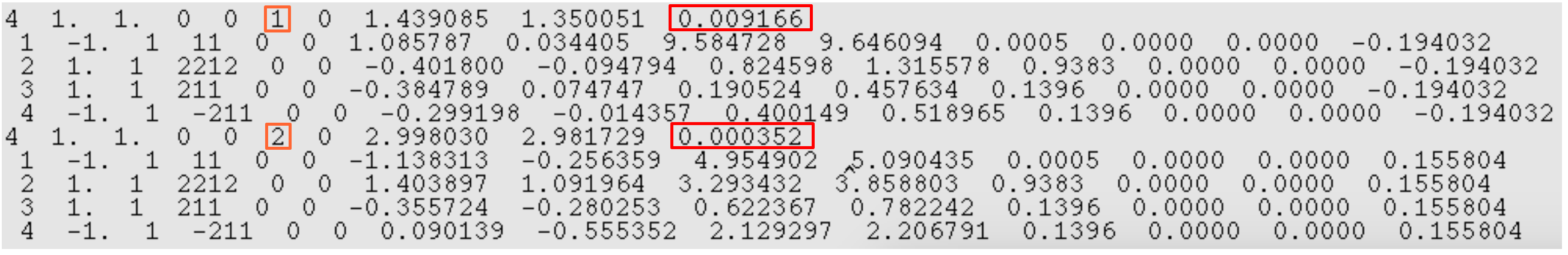}
\end{center}
\caption{\small Example of two generated double pion events. The event numbers are in the orange boxes, while the weights are in the red boxes.}
\label{fig:evnt_exmpl}
\end{figure}

The generator produces output in the LUND format. The header of the event was slightly changed in comparison with the conventional one that is used in GEMC. In the field six the event number is placed instead of "x" and in the field ten the cross section value is placed instead of "nu"  (see Fig.~\ref{fig:lund_out}). An example of LUND output for two double pion events generated with TWOPEG is given in Fig.~\ref{fig:evnt_exmpl}.

For the backward compatibility with CLAS software TWOPEG can also produce  output in BOS format. In this case there are two options in the input file: the option with $F_{bos}=1$ serves to generate "MCTK" and "MCVX" banks, while the other option with $F_{bos}=2$ is needed to generate the "PART" bank.

TWOPEG needs ".dat" files with tabulated structure functions and fit parameters. They are located in the "data" subfolder inside the EG root directory. If one needs to move it, one should define the environment variable "data\_dir\_2pi" that points to the new "data" folder location (for example in csh one should use "setenv data\_dir\_2pi new\_path/").

There are also two ".root" output files: the first one contains the root tree with the weights for each event and the second one contains some histograms, which one may find useful for the purpose of a quick check of the kinematical coverage, distributions of EG yield, etc.

\chapter{Conclusions and code availability}
\label{sect:concl}

TWOPEG code is available in the github repository (https://github.com/JeffersonLab/Hybrid-Baryons/). 
One can obtain the code using the command: \\
 {\em "git clone  https://github.com/JeffersonLab/Hybrid-Baryons.git"}.

TWOPEG covers the kinematical region in $W$ from the double pion production threshold up to 4.5~GeV. For $W > 2.5$~GeV the higher the value of $W$ is, the less reliable the cross section shape becomes, since neither experimental data nor model predictions exist there.

In $Q^2$ the EG works starting from 0.0005~GeV$^2$. In general it can work for any $Q^2$ greater than this lower limit, but one should remember that for $Q^2 > 1.3$~GeV$^2$ the shape of the differential cross sections is always the same as for $Q^2 = 1.3$~GeV$^2$ and only integral scaling takes place.

In the kinematical regions, where the information about the cross sections of the reaction $e p \rightarrow e' p' \pi^+ \pi^-$ exists (regions 1a, 1b, and 2a in Fig.~\ref{fig:gen_cover}) TWOPEG successfully reproduces them. In the regions I, II, and IV in Fig.~\ref{fig:gen_cover} the EG predicts the cross section value based on the known cross sections in the neighboring regions, therefore it can be used as a naive model there. In other regions the cross section estimation is less reliable, but good enough for the purpose of the modeling of experiments, for example for efficiency evaluation or background estimation.

TWOPEG was developed in the framework of the preparation of the Hybrid Baryon Search proposal, which was approved by PAC44~\cite{PAC44:Hybrid}.
It has already been successfully used for the run condition and efficiency estimations
during the proposal preparation.
For the event reconstruction the simplified version of the CLAS12 reconstruction software (FASTMC) was employed.
TWOPEG will be applied for the two pion analysis of the CLAS12 data that are expected for this project.  

It also will be used for the analysis of the CLAS12 data that will be collected in connection with the Nucleon Resonance Studies With CLAS12 proposal approved by PAC34~\cite{PAC34:twopi}.

The cross sections obtained from TWOPEG have already been used to fill zones with zero CLAS acceptance in the analysis of the part of "e1e" data-set that ran with the hydrogen target~\cite{Fedotov:note}.

In the analysis of the part of "e1e" data-set with deuteron target 
the efficiency evaluation and the corrections due to the radiative effects and Fermi motion of the target proton have been carried out with TWOPEG~\cite{Skorodum:note}. For that purpose the generated events were passed through the standard CLAS packages "GSIM" and "recsis".

TWOPEG is being established as a universal tool for Monte Carlo simulation of the reaction $e p \rightarrow e' p' \pi^+ \pi^-$ and can be used for future CLAS12 experiments as well as in the continuing CLAS data analyses.

\appendix
\renewcommand{\thechapter}{A}
 \refstepcounter{chapter}
    \makeatletter
   \renewcommand{\theequation}{\thechapter.\@arabic\c@equation}
    \makeatother
\chapter*{\LARGE Appendix A: Calculation of the angle $\alpha$}
\label{app_a}
\addcontentsline{toc}{chapter}{Appendix A: Calculation of the angle~$\alpha$}

For the second set of kinematic variables the angle $\alpha_{\pi^{-}}$ between two planes A and B (see Fig.~\ref{fig:alpha_2nd_set}) should be calculated in the following way. Firstly two
auxiliary vectors $\vec \gamma$  and
$\vec \beta$ should be determined. The vector $\vec \gamma$ is the unit vector perpendicular to the three-momentum
$\vec P_{\pi^{-}}$, directed toward the vector $(-\vec n_{z})$ and situated in the plane A, which is defined by 
the three-momentum of initial proton and three-momentum of $\pi^{-}$. $\vec
n_{z}$ is the unit vector directed along $z$-axis.
The vector $\vec \beta$ is the unit vector perpendicular to the three-momentum of $\pi^{-}$, 
directed toward the three-momentum of $\pi^{+}$ and situated in the plane B, which is defined 
by all final hadrons. Note that the three-momenta of $\pi^{+}$,
$\pi^{-}$, and $p'$ are in the same plane, since in c.m. frame
their total three-momentum has to be equal to zero.
 Then the angle between two planes  $\alpha_{\pi^{-}}$ is
\begin{equation}
\alpha_{\pi^{-}} = acos(\vec \gamma \cdot \vec \beta),
\label{eq:cr_sec_anglealpha}
\end{equation}
where $acos$ is a function that runs between zero and
$\pi$, while the angle $\alpha_{\pi^{-}}$ may vary between zero and
$2\pi$. To determine the $\alpha$ angle in the
range between $\pi$ and $2\pi$ 
the relative direction between the $\pi^{-}$ three-momentum and the vector product $\vec \delta = [ \vec \gamma \times \vec \beta ]$ of the auxiliary vectors $\vec
\gamma$ and $\vec \beta$ should be taken into account.
If the vector $\vec \delta$ is collinear to the three-momentum of $\pi^{-}$, the angle $\alpha_{\pi^{-}}$ is determined
by~(\ref{eq:cr_sec_anglealpha}), and in a case of anti-collinearity by
\begin{equation}
\alpha_{\pi^{-}} = 2\pi - acos(\vec \gamma \cdot \vec \beta).
\label{eq:cr_sec_anglealpha_var}
\end{equation}
The defined above vector $\vec \gamma$ can be expressed as
\begin{eqnarray}
\vec \gamma = a_{\alpha}(-\vec n_{z}) + b_{\alpha}\vec n_{P_{\pi^{-}}} & \text{with} \nonumber \\
a_{\alpha} = \sqrt{\frac{1}{1 - (\vec n_{P_{\pi^{-}}} \cdot (-\vec n_{z} ) )^{2}}} & \text{and} \label{alphavec}\\
b_{\alpha} = - (\vec n_{P_{\pi^{-}}} \cdot (-\vec n_{z} ) ) a_{\alpha} \textrm{ ,} \nonumber
\end{eqnarray} 
where $\vec n_{P_{\pi^{-}}}$ is the unit vector directed along the three-momentum of $\pi^{-}$ (see Fig.~\ref{fig:alpha_2nd_set}).

Taking the scalar products $(\vec \gamma \cdot \vec
n_{P_{\pi^{-}}})$ and $(\vec \gamma \cdot \vec  \gamma)$,
it is straightforward to verify, that $\vec \gamma$ is the unit vector perpendicular to the three-momentum of $\pi^{-}$.

The vector $\vec \beta$ can be obtained as
\begin{eqnarray}
\vec \beta = a_{\beta}\vec n_{P_{\pi^{+}}} + b_{\beta}\vec n_{P_{\pi^{-}}} & \text{with} \nonumber \\
a_{\beta} = \sqrt{\frac{1}{1 - (\vec n_{P_{\pi^{+}}} \cdot \vec n_{P_{\pi^{-}}})^{2}}} & \text{and} \label{betavec}\\
b_{\beta} = - (\vec n_{P_{\pi^{+}}} \cdot \vec n_{P_{\pi^{-}}}) a_{\beta} \textrm{ ,} \nonumber
\end{eqnarray} 
where $\vec n_{P_{\pi^{+}}}$ is the unit vector directed along the three-momentum of $\pi^{+}$.

Again taking the scalar products $(\vec \beta \cdot \vec
n_{P_{\pi^{-}}})$ and $(\vec \beta \cdot \vec  \beta)$,
it is straightforward to see, that $\vec \beta$ is
the unit vector perpendicular to the 
three-momentum of $\pi^{-}$. 

The angle $\alpha_{\pi^{-}}$ coincides with
 the angle between the vectors $\vec \gamma$ and
$\vec \beta$.
So, the scalar product $( \vec \gamma \cdot
\vec \beta )$ allows to determine the angle
$\alpha_{\pi^{-}}$~(\ref{eq:cr_sec_anglealpha}). The angles $\alpha_{p'}$ and $\alpha_{\pi^+}$ from the other sets of kinematic
variables are calculated in the similar
way.


\bibliography{gen_note}

\newpage
\end{document}